%% file: main.tex
\documentclass[letterpaper,twocolumn,prb,superscriptaddress,longbibliography]{revtex4-2}
\include{header}

\usepackage[normalem]{ulem}

\begin{document}

\title{
Dimensional reduction of Kitaev spin liquid at quantum criticality
}
\author{Shi Feng}
\email[E-mail:]{feng.934@osu.edu}
\affiliation{Department of Physics, The Ohio State University, Columbus, Ohio 43210, USA}
\author{Adhip Agarwala}
\email[E-mail:]{adhip@iitk.ac.in}
\affiliation{Department of Physics, Indian Institute of Technology, Kanpur 208016, India}
\author{Nandini Trivedi}
\email[E-mail:]{trivedi.15@osu.edu}
\affiliation{Department of Physics, The Ohio State University, Columbus, Ohio 43210, USA}

\begin{abstract}
We investigate the fate of the Kitaev spin liquid (KSL) under the influence of an external magnetic field $h$ in the [001] direction and upon tuning bond anisotropy of the Kitaev coupling $K_z$ keeping $K_x = K_y = K$. 
Guided by density matrix renormalization group, exact diagonalization, and with insights from parton mean field theory, we uncover a field-induced gapless-to-gapless Lifshitz transition from the nodal KSL to an intermediate gapless phase. 
The intermediate phase sandwiched between $h_{c1}$ and $h_{c2}$, which persists for a wide range of anisotropy $K_z/K > 0$, is composed of 
weakly coupled one-dimensional quantum critical chains. This intermediate phase is a dimensional crossover which asymptotically leads to the one-dimensional quantum Ising criticality characterized by the (1+1)D conformal field theory as the field reaches
the phase transition at $h_{c2}$. 
Beyond $h_{c2}$ the system enters a partially polarized phase describable as effectively decoupled bosonic chains in which spin waves propagate along the one-dimensional zigzag direction. 
Our findings provide a comprehensive phase diagram and offer insights into the unusual physics of dimensional reduction generated by a uniform magnetic field in an otherwise two-dimensional quantum spin liquid.

\end{abstract}
\date{\today}
\maketitle
\section{Introduction}
\begin{figure}[h]
    \centering
    \includegraphics[width=\linewidth]{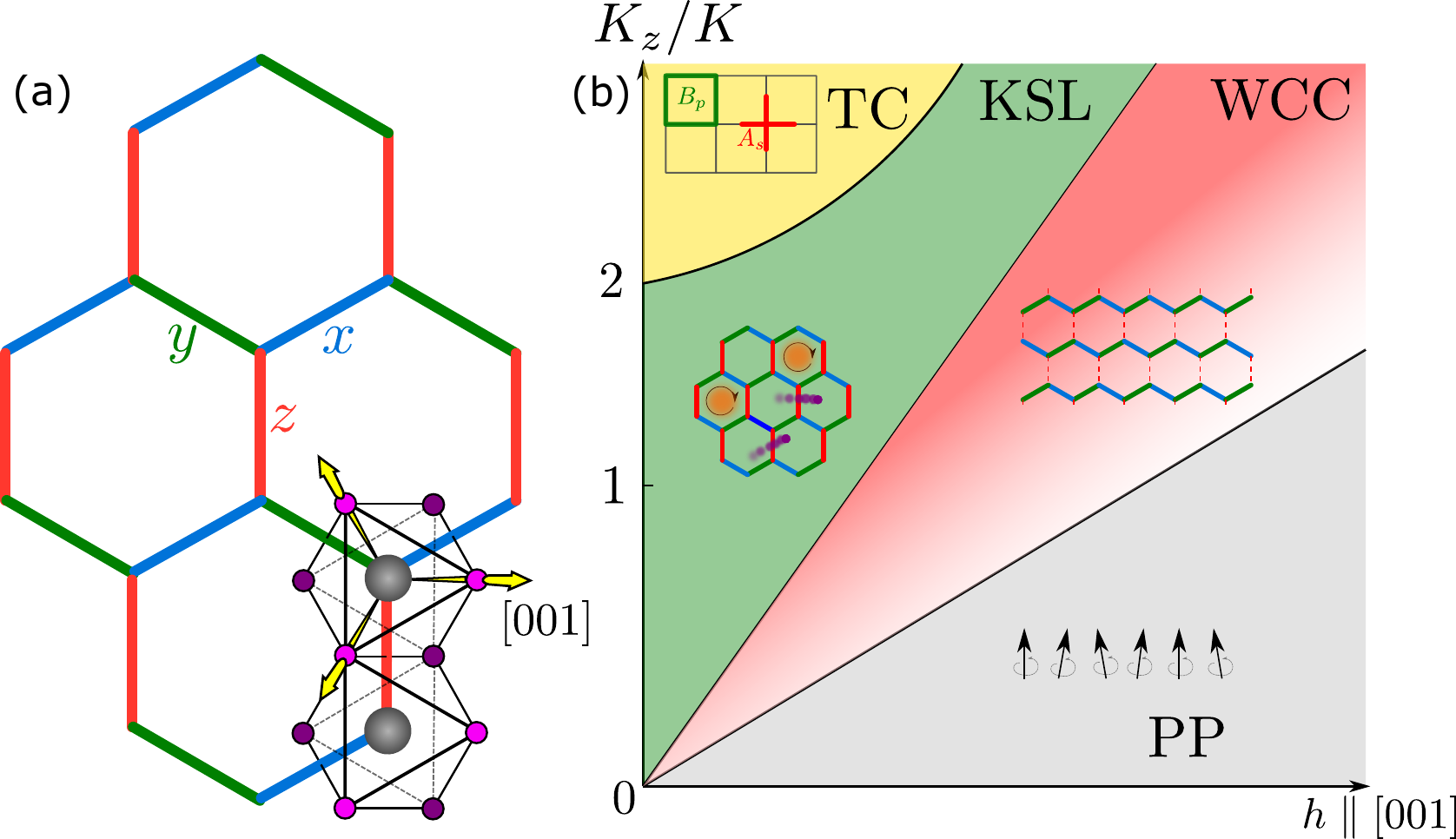}
    \caption{(a) Bond dependent spin exchange interactions in the Kitaev honeycomb model.
    The lower-right corner illustrates the edge-sharing octahedra of $\alpha$-RuCl$_3$, whereas the grey bullets on the vertices denote Ru ions, and the violet and pink circles surrounding them are ligands of Cl above and below the honeycomb plane. 
    The [001] field points from the ion to one of its ligands. (b) The schematic phase diagram as a function of $(K_z,h)$. The inset in each phase illustrates the phase: Toric code (TC), Kitaev spin liquid (KSL), weakly-coupled chains (WCC) and the partially polarized (PP) phase. The red-to-white color gradient illustrates the WCC as the crossover from 2D physics due to the non-zero inter-chain coupling, to the 1D physics where the inter-chain coupling asymptotically vanishes as the field reaches quantum the critical point at $h_{c2}$, confining excitations across zigzag chains. }
    \label{schematic}
\end{figure}

The Kitaev model is a paradigmatic model for an exactly solvable quantum spin liquid (QSL), consisting of spin S$=\frac{1}{2}$ local magnetic moments on a two-dimensional honeycomb lattice with compass-like bond-dependent interactions given by the Hamiltonian $H_K = \sum_{\langle i,j\rangle, \alpha}{K_\alpha \sigma^\alpha_i \sigma^\alpha_j}$ \cite{kitaev2006anyons}, where $\alpha=x,y,z$ labels the three bonds of the honeycomb lattice and the nature of Ising interactions thereof. The low energy excitations of the QSL are the Majorana fermions with a two-dimensional Dirac like dispersion. It is natural to pose if such a two dimensional QSL phase can be assembled from  effective one-dimensional phases. In a manner analogous to the wire construction of the quantum Hall state, where 1D chiral edge modes are interwoven to form a 2D topologically ordered state \cite{Kane2002,Kane2014,Marcel2020}, the Kitaev model can indeed be visualized as an assembly of XY compass chains, that is, the zigzag chains made up of blue and green bonds in Fig. \ref{schematic}(a), interconnected by spin exchanges along the $\hat{z}$ axis, which corresponds to the nature of interaction on the red-colored $z$ bonds in Fig. \ref{schematic}(a). The fluxes inherent to the Kitaev solution can be perceived as the manifestation of the flux quantum in a quantum Hall system due to the coupling of free fermions between adjacent chiral fermionic chains. Notably, this picture has recently been explored theoretically to show that the Kitaev honeycomb model can be derived from quantum Ising critical chains via fine-tuned interchain couplings \cite{Mudry2017,Slagle2022,liu2023}.

In this study, we pose the opposite question: Can the Kitaev spin liquid be effectively transformed into decouped quantum Ising chains via a non-fine-tuned and robust external parameter?
As we will show, a uniform magnetic field in either $x,y$ or $z$ directions, presents a natural choice for such an external parameter. This would be analogous to the field-induced dimensional crossover from a 3D frustrated insulator to an effective 2D model near the quantum critical point \cite{Fisher2006,Fisher2007,Batista2008,Zenji2021}, but in a microscopic spin model with one lower dimension. Previous experimental studies have already shown intriguing results in 2D Mott insulators related to the field-induced quantum criticality. For instance, a field-induced quantum critical point has been observed in $\alpha$-RuCl$_3$ \cite{Wolter2017,Nagai2020} which may be proximate to the Kitaev quantum spin liquid state. The impact of such a field on the Kitaev spin liquid has also been extensively researched theoretically, establishing the existence of a gapless intermediate phase in the Kitaev model under a Zeeman magnetic field applied in various directions \cite{Gohlke_PRB_2018,David2019,Patel12199,hickey2019emergence,Hickey20212,Hickey20221,Hwang2022,Liang2018,Nasu2018,Feng20222}. Following these developments, in this work we demonstrate that the effect of a uniform magnetic field along the [001], [010], or [100] axes leads to an effective dimensional reduction from the two-dimensional Dirac spin liquid to decoupled Ising critical chains, which is in sharp contrast to the anti-ferromagnetic Heisenberg model on a honeycomb lattice where no dimensional reduction is present under arbitrary Zeeman field due to the U(1) symmetry \cite{Fransson2016}.



For completeness, we first present the phase diagram generated by a magnetic field along the [001] direction and bond anisotropy $K_z$ and contrast that with the phase diagram for a field along [111]~\cite{Yuanming2018,Feng20222}. In the latter case, there is an intermediate gapless phase, however its extent in the phase diagram is significantly smaller. We also find the emergence of one-dimensional quantum Ising criticality within the intermediate phase. This criticality approaches an asymptotically exact description as a critical transverse field Ising model (TFIM) near the phase boundary at $h_{c2}$. Given the recent experiments on $\alpha$-RuCl$_3$ demonstrating field-induced quantum criticality under the influence of a magnetic field \cite{Wolter2017,Nagai2020}, our results offer a theoretical underpinning for these observations.

The Hamiltonian of interest is given by the Kitaev model under a [001] Zeeman field:
\begin{equation}
H = \sum_{i, \alpha} K_{\alpha} \Big(  \sigma^\alpha_{i,A}  \sigma^\alpha_{i,A+\hat{\alpha}}  \Big) - \sum_{i} h (\sigma^{z}_{i,A} + \sigma^{z}_{i,A+\hat{z}})
\label{eq_spinHam}
\end{equation}
with antiferromagnetic exchange $K_\alpha > 0$ and $\alpha \in \{x,y,z\}$. $A(B)$ label different sublattices, and the subscript $A + \hat{\alpha}$ is equivalent to the $B$ sublattice of the same or another unit cell separated by a translation along the $\alpha$ bond.
We further investigate the effect of a [001] field and the exchange anisotropy $K_z/K$ ($K_x = K_y \equiv K$) on the Kitaev model using exact diagonalization (ED) and density matrix renormalization group (DMRG)techniques. We gain further insight into our results from majorana mean field theory (MFT) and linear spin wave (LSM) analysis in the high field regime. The central result of our work is that the gapless phase induced by the [001] field is effectively a quasi-one-dimensional fermionic model, i.e. an assembly of \emph{weakly coupled chains} (WCC). Its low energy description should be captured by a (1+1)D conformal field theory (CFT) with central charge $c=\frac{1}{2}$ and weakly coupled left and right chiral majoranas; and the partially polarized (PP) phase is effectively \emph{decoupled} bosonic chains. Figure~\ref{schematic} shows the setup of the problem and the schematic phase diagram of the Kitaev model as a function of ${K_z, h}$.

\section{Phase diagram}
Figure \ref{schematic}(b) summarizes all the phases induced by the anisotropy of exchange and [001] field. At small field and low anisotropy, the Hamiltonian results in the gapless Kitaev spin liquid whose elementary excitations are itinerant majorana fermions and $Z_2$ fluxes. Note that in contrast to the [111] field, [001] field doesn't open a gap in the majorana sector. At small [001] field, the KSL remains a gapless phase whose spin-spin correlation decays by a power law \cite{Kitaev2011}. At small field and high anisotropy, the model enters the Toric code (TC) phase, whose effective degrees of freedoms are given by $z$ dimers with the low energy manifold $\{\ket{\up\dn},~\ket{\dn\up}\}$ \cite{Nanda_PRB_2021}. At the intermediate field proportional to $K_z/K$, the Hamiltonian gives the intermediate WCC phase, as an effective model which features weakly coupled fermionic chains, i.e. the WCC phase is continuously connected to the $(K_z=0, h=0)$ point as decoupled 1D compass chains, which we discuss below.  Notably, unlike the intermediate gapless phase induced by the [111] field which terminates at a finite $K_z/K$ \cite{Yuanming2018,Feng20222}, the gapless phase induced here by the [001] field persists up to considerably larger anisotropy, at least up to $K_z/K \sim 3$, as is shown schematically in Fig.~\ref{schematic}(b).  

\begin{figure}[t]
    \centering
    \includegraphics[width=\linewidth]{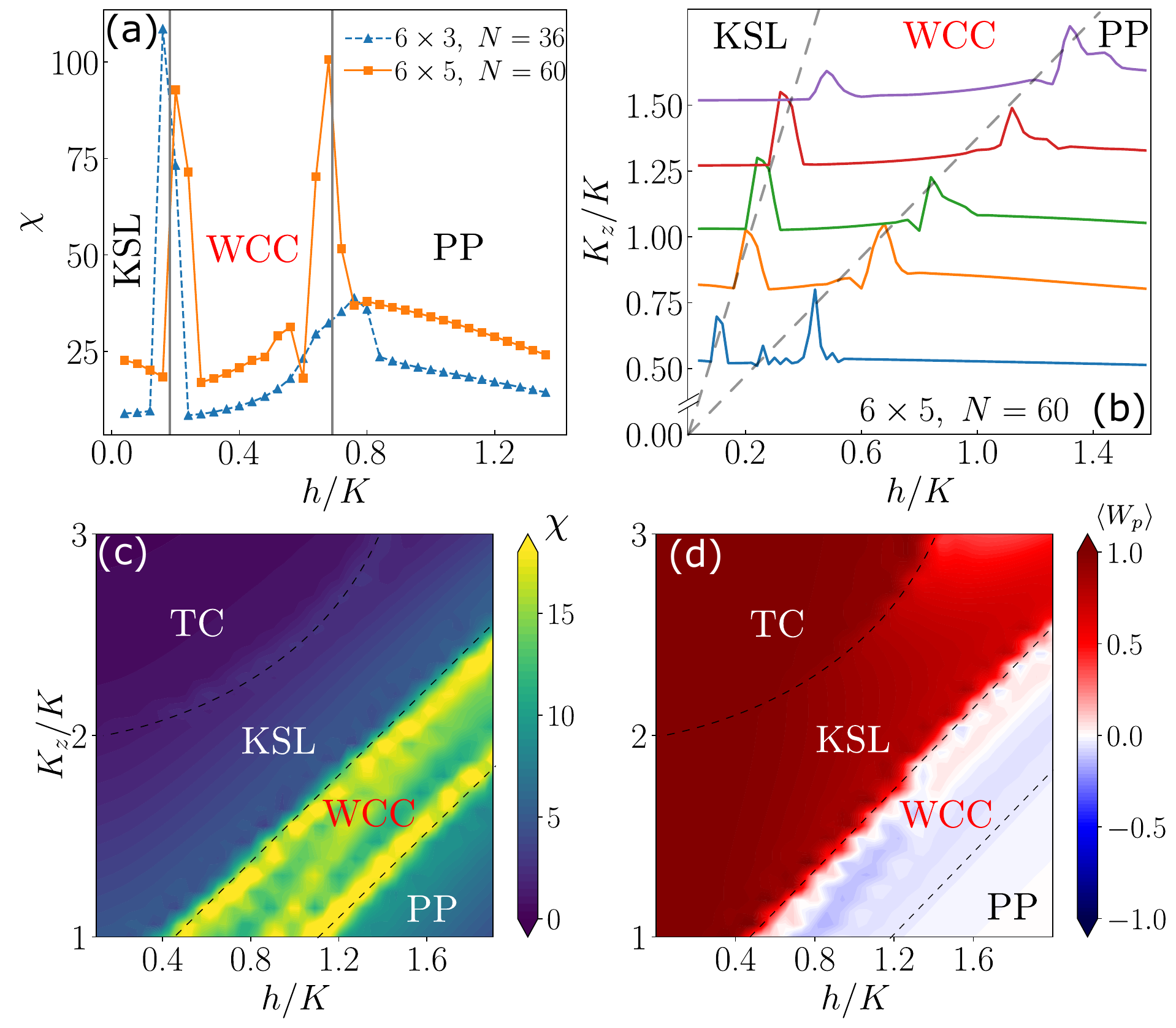}
    \caption{Diagnostics of phases using magnetic susceptibility $\chi$ and plaquette fluxes $\expval{W_p}$. (a) $\chi$ as a function of $h$ at $K_z/K = 1$. (b) Scaled cuts of $\chi$ in the $(K_z, h)$ plane in a.u. ;  the black dashed lines identify the divergences of $\chi$ that marks out an expanding critical region as the anisotropy increases. (c) Density plot of $\chi$ in the $(K_z, h)$ plane. (d) $\expval{W_p}$ as a function of $(K_z, h)$. Data in (a) and (b) are obtained on 36-site ($6\times 3$ unit cells) and 60-site ($6\times 5$ unit cells) using DMRG with OBC. Data in (c) and (d) are obtained by 24-site ($3\times 4$ unit cells) ED with PBC.}
    \label{EDDMRG}
\end{figure}

\begin{figure*}
    \centering
    \includegraphics[width=\linewidth]{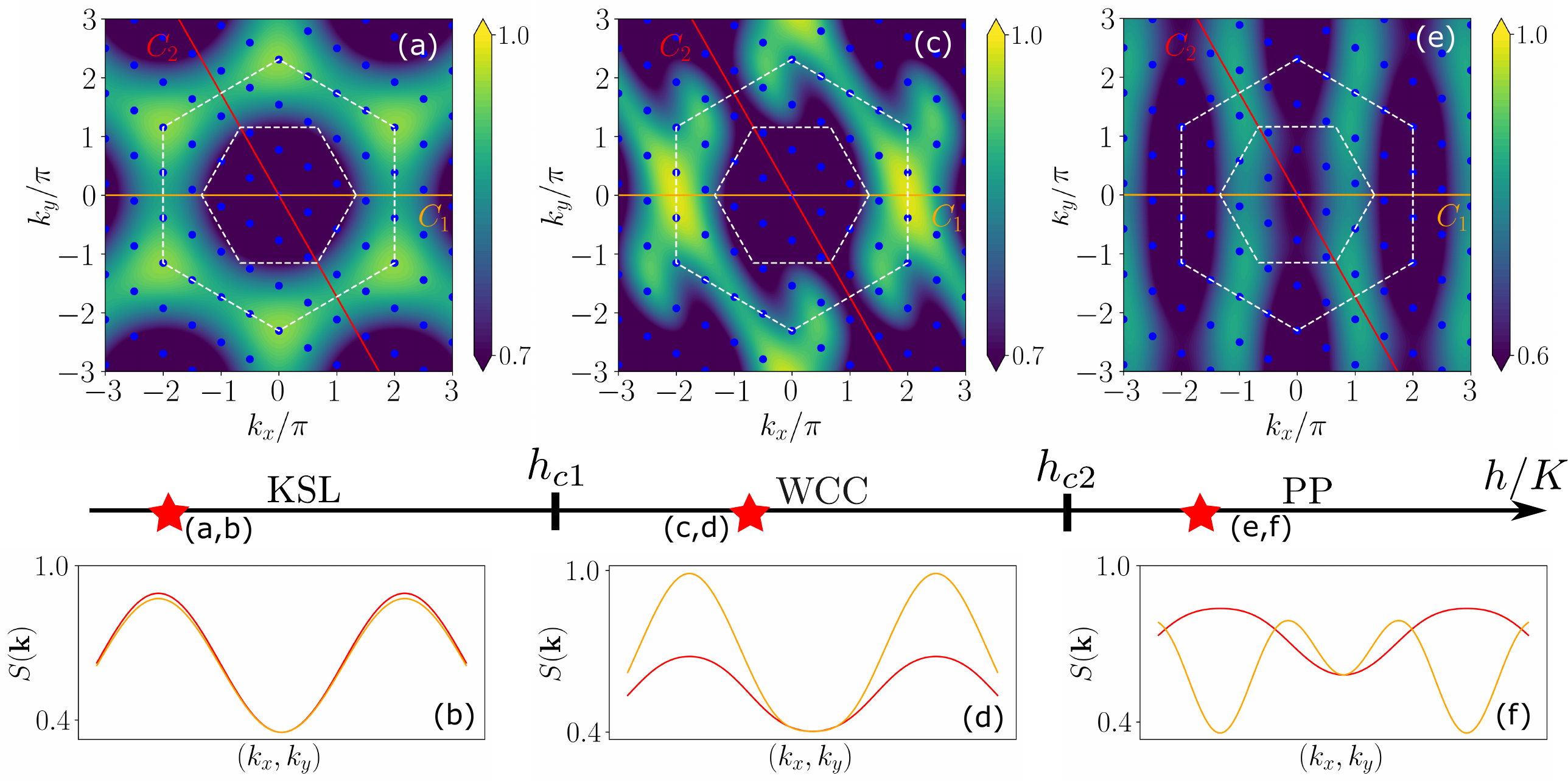}
    \caption{The ground-state static structure factor $S(\mathbf{k})$, as defined in Eq. \ref{eq:ss}, in three different phases.  White dashed lines mark the first and extended Brillouin zones. (a,b) $S(\mathbf{k})$ at $(K_z/K = 1, h/K = 0.04)$ in KSL phase along cuts indicated by the orange ($C_1$) and red ($C_2$) lines. (c,d) $S(\mathbf{k})$ at $(K_z/K = 1, h/K=0.7)$ in WCC phase. (e,f) $S(\mathbf{k})$ at $(K_z/K = 1, h/K=1.4)$ in the PP phase.  Data obtained by $24$-site ED. The blue dots denote the momentum space resolution of the 24-site cluster.}
    \label{fig3}
\end{figure*}

We diagnose these phases by ED on 24-site clusters with periodic boundary conditions (PBC) and using DMRG calculations under open boundary conditions (OBC) \cite{itensor,itensor-r0.3} (See also Appendix \ref{app:finite}), and complement these with a mean field analysis. The observables we monitor are the magnetic susceptibility $\chi$ and the expectation value of the flux  $\expval{W_p}$ in a hexagonal plaquette:
\begin{equation}
    \chi = \frac{\partial^2 E_{\text{gs}}(K_z,h)}{\partial h^2},~
    \expval{W_p} = \expval{\sigma_1^x \sigma_2^y \sigma_3^z \sigma_4^x \sigma_5^y \sigma_6^z}
\end{equation}
$\chi$ is sensitive to diverging correlation lengths that arise close to phase transitions, and 
$ \expval{W_p}$ is an indicator of emergent $Z_2$ gauge fields. 
Figure~\ref{EDDMRG}(a) shows the magnetic susceptibility $\chi$ as a function of $h$ at fixed $K_z/K = 1$. The two diverging peaks at $h_{c1}$ and $h_{c2}$ mark out the transition from KSL to WCC, and from WCC to PP. The data is obtained using DMRG on a cylindrical geometry. Fig.~\ref{EDDMRG}(b) shows $\chi$ with different values of $K_z$ anisotropy, which marks out the two phase boundaries previously sketched in Fig.~\ref{schematic}(b). It is also supported by ED calculation, as is shown in Fig.~\ref{EDDMRG}(c): there are obviously four distinct regions in the contour shown in both panels marked out by singularities; in particular, the divergence of the susceptibilities in the intermediate field region and low anisotropy is indicative of a gapless phase whereby $\xi$ diverges, which is consistent with previous works \cite{Liang2018,Nasu2018} and our MFT (Appendix \ref{sec:MFT}). 
The average flux calculated in the same parameter scan, shown in Fig. \ref{EDDMRG}(d), also supports the phase diagram.  
Note the average flux $\expval{W_p}$ becomes negative in the WCC phase, indicating strong quantum fluctuations in the gauge sector with the proliferation of plaquette fluxes, in contrast to the PP where $\expval{W_p} \approx 0$ due to the confinement of partons. This is similar to the [111]-induced gapless phase, where a putative ``glassiness" of high-density fluxes in the gauge sector has been recently reported \cite{yogendra2023emergent}. We will revisit the gauge fluctuations in the parton mean field analysis in the forthcoming sections. 

\section{Scattering signatures}
In this section we discuss signatures relevant for scattering experiments in KSL, WCC and PP phases, including both static and dynamical structure factors; as well as high-field limit linear spin wave theory. For consistency, all numerically obtained quantities in this section are obtained by 24-site ED under PBC.  Larger scale DMRG calculation under OBC is discussed in Appendix \ref{app:finite}, and is consistent with ED calculation. 
\subsection{Signature of WCC in equilibrium}
The magnetic field along [001] explicitly breaks the mirror symmetry along $x$ or $y$ bond, hence it is necessary to investigate the correlations along $z$ bonds and along $x/y$ bonds separately. 
This is achieved by calculating the resolution of the total magnetic fluctuations in momentum space:
\begin{equation} \label{eq:ss}
    S(\mathbf{k}) = \expval{\boldsymbol{\sigma}(\mathbf{k})\cdot \boldsymbol{\sigma}(-\mathbf{k})} - \expval{\boldsymbol{\sigma}(\mathbf{k})}\cdot \expval{\boldsymbol{\sigma}(-\mathbf{k})}
\end{equation}
for KSL and WCC phases using the $24$-site ED, shown in Fig. \ref{fig3}. In KSL, the dominant correlation is from the nearest neighbor $\expval*{\sigma_i^\alpha \sigma_{i+\alpha}^\alpha}$ for $\alpha = x,y,z$ \cite{Baskaran2007}, therefore, $S(\mathbf{k})$ has its strongest intensities around the extended Brillouin zone boundary as shown in Fig.~\ref{fig3}(a,b). The pattern of $S(\mathbf{k})$ changes dramatically in the WCC phase when $h_{c1}<h<h_{c2}$ where the three-fold rotational symmetry is broken, as shown in Fig. \ref{fig3}(c,d). The dominant signal exhibits a stripy pattern parallel to $k_y$ near the second Brillouin zone, along with weaker noisy signals outside the stripes that reflect the existence of weak coupling between zigzag chains. Hence the quantum fluctuation in the WCC phase is predominantly along the 1D zigzag chains, and the weak coupling between chains diminishes as $h \rightarrow h_{c2}^-$. Notably, the $\expval*{\delta \sigma_i^z \delta \sigma_{i+z}^z}$ channel in Eq. \ref{eq:ss} becomes negligible in the WCC phase, as the Zeeman field suppresses quantum fluctuations along the field direction. This is similar to the intermediate gapless phase induced by the [111] field, where the out-of-plane fluctuation becomes negligible while leaving most of the fluctuation within the plane \cite{zhang2023machine}. With larger field $h>h_{c2}$, the system becomes polarized along [001], whose ground state fluctuation is also strongly anisotropic as shown in Fig.~\ref{fig3}(e,f). We will revisit the quasi-1D nature of the bosonic excitation in PP in the next section.

\begin{figure}
    \centering
    \includegraphics[width=0.99\linewidth]{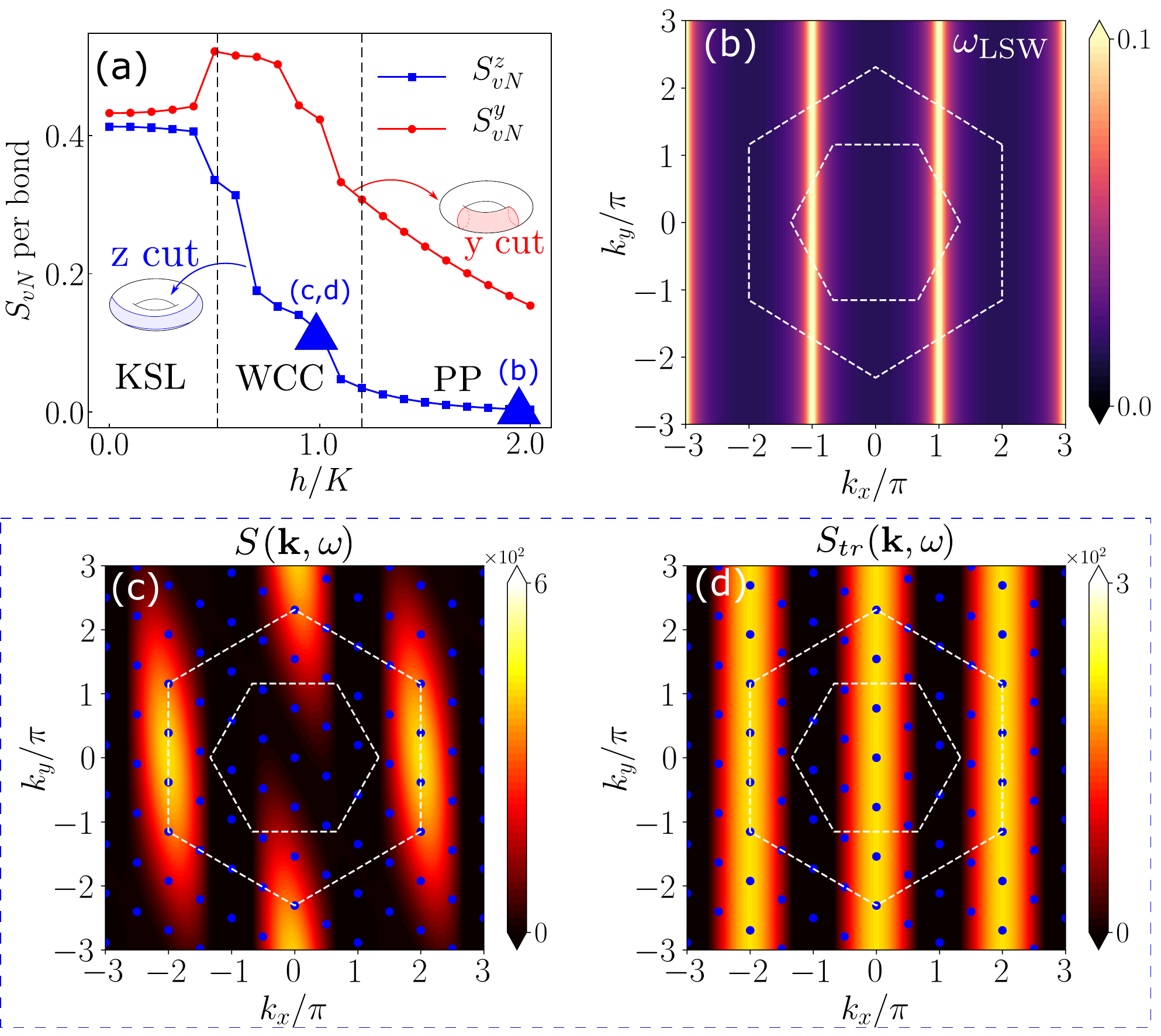}
    \caption{(a) von-Neumann entanglement entropy per bond $S_{\rm vN}$ of the subsystem enclosed by two Wilson loops that cut through six z bonds in total (blue); and $S_{\rm vN}$ of the subsystem enclosed by two Wilson loops that cut through eight y bonds in total (red), as is illustrated in the inset torus. 
    (b) The lower magnon band of the PP phase obtained by linear spin wave (LSW) theory. 
    (c) the total dynamical structure factor $S(\mathbf{k},\omega)$, and (d) the intra-sublattice dynamical structure factor $S_{tr}(\mathbf{k},\omega)$ at $\omega = 0.018$ for $h/K = 1.0$ $(< h_{c2}^{\rm ED}/K \approx 1.1)$ inside the WCC phase. (a,c,d) are obtained by 24-site ED at different $h$ with $K_z/K = 1$. The blue dots denote the momentum space resolution of the cluster.}
    \label{fig:dyn}
\end{figure}

The above characterization of phases is further supported by the von-Neumann entanglement entropy of subsystems whose boundaries cut through different sets of bonds. Knowing the ground state density matrix $\rho$, the von-Neumann entropy $S_{\rm vN}$ is obtained by summing over the eigenvalues of the reduced density matrix $\rho_A$ of subsystem $A$: $S_{\rm vN}(A) = -\Tr \rho_A \log \rho_A$ with $\rho_A = \Tr_{\bar{A}} \rho$, where $\bar{A}$ is the complement of $A$. The calculated $S_{\rm vN}$ per bond for subsystems whose boundaries cut through $z$ and $y$ bonds respectively are shown in Fig.~\ref{fig:dyn}(a) and Appendix~\ref{sec:suppres}. It clearly detects the highly anisotropic entanglement in the WCC phase in which the entanglement entropy per $z$ bond is very small $S_{\rm vN}^z({\rm WCC}) \sim 0.1$, while that for a $y$ bond cut is much stronger, with $S_{\rm vN}^y({\rm WCC})\sim 0.5$ which is close to the maximal Bell pair entanglement of $\ln 2$.  In contrast, the mirror symmetry in the KSL phase persists and the entanglement entropy remains $S_{\rm vN}^z({\rm KSL}) \approx S_{\rm vN}^y({\rm KSL})$. Note that at the integrable limit the correlation is extremely short-ranged, hence the entanglement per bond can be easily estimated to be $S_{\rm vN}^{z(y)}({\rm KSL}) \sim 0.5$ \cite{Feng2022pra}, which is close to the ED result for KSL shown in Fig. \ref{fig:dyn}(a). The weak entanglement along $z$ bonds in WCC phase compared to that in KSL indicates the effective exchange coupling along $z$ bonds is strongly suppressed by the [001] field, while the coupling along $x$ and $y$ bonds are relatively enhanced. Therefore the WCC phase is a lot more coherent within the zigzag chains than across the zigzag chains.
This also supports the results obtained by majorana mean field theory \cite{Liang2018,Nasu2018} where majoranas under the mean field exhibit quasi-one dimensional dispersion in the intermediate phase. We will later elaborate on the MFT results and investigate the effective spin exchange induced by the field. 
Interestingly, emergent 1D behavior is also present in the spin wave dispersion of PP phase, where the energy density of $z$ bonds in the effective LSW Hamiltonian vanishes as $h \rightarrow h_{c2}^+$ in the PP phase, resulting in 1D magnon modes propagating along zigzag chains only, as shown in Fig. \ref{fig:dyn}(b) and further details in Appendix \ref{sec:lsw}. 
Indeed, one may also understand the PP in a fermionic picture according to \cite{Sun2009}: If we start from the decoupled fermionic chain picture at $h_{c2}$, a non-zero field in $\hat{z}$ direction immediately gaps out the free fermion; and the inter-chain coupling $\sigma_i^z \sigma_{i+z}^z$ at $h>h_{c2}$ corresponds only to an interaction between fermion occupation numbers that share a $z$-type bond, giving no contribution of inter-chain hopping. Hence at high field the only non-trivial dynamics is due to the perturbation of intra-chain hopping terms.

\subsection{Low energy dynamics of WCC}
Here we investigate the leading order dynamics of the WCC phase by ED complementary to previous DMRG data. It can be difficult to accurately capture the dynamics within DMRG for gapless states. We will show that the intermediate gapless phase exhibits emergent one-dimensional dynamics despite inter-chain interactions. 
Given that the unit cell of the honeycomb lattice has two sites that belong to sublattices $A$ and $B$, we define two different dynamical structure factors \cite{Patel12199}: (1) The total $S(\mathbf{k},\omega)$ as the Fourier transform of all sites irrespective of the sublattice index, which, similar to the static structure factor in Eq. \ref{eq:ss}, is periodic within the second Brillouin zone, and up to a normalizing factor is defined by
\begin{equation}
    S^{\alpha\beta}(\mathbf{k},\omega) \propto  \mathfrak{I} \left[ \sum_{i,j} \langle \sigma_{\mathbf{r}_i}^\alpha \frac{1}{\omega - H + i \epsilon} \sigma^\beta_{\mathbf{r}_j} \rangle e^{i\mathbf{k}\cdot\mathbf{r}_{ij}}\right]
\end{equation}
where $\mathbf{r}_{ij} = \mathbf{r}_i - \mathbf{r}_j$; 
and (2) The ``traced" dynamical structure factor $S_{tr}(\mathbf{k},\omega)$ that includes only intra-sublattice spin-spin correlations, i.e. correlations on the Bravais lattice, and is therefore periodic within the first Brillouin zone, defined by
\begin{equation}
    S_{tr}^{\alpha\beta}(\mathbf{k},\omega) \propto  \mathfrak{I} \left[  \sum_\kappa \sum_{ (i,j) \in \kappa} \langle \sigma^\alpha_{\mathbf{r}_i} \frac{1}{\omega - H + i \epsilon} \sigma^\beta_{\mathbf{r}_j} \rangle e^{i\mathbf{k}\cdot\mathbf{r}_{ij}}\right]
\end{equation}
 where $\epsilon \sim 0^-$ is a small broadening factor which we set as $10^{-3}$ in calculation, $\kappa \in \{A,B\}$ is sublattice index, and $\mathfrak{J}$ takes the imaginary part of the expression.  

We calculate the symmetric dynamical structure factors $S(\mathbf{k},\omega) = \sum_{\gamma\in\{x,y,z\}} S^{\gamma\gamma}(\mathbf{k},\omega)$ and $S_{tr}(\mathbf{k},\omega) = \sum_{\gamma\in\{x,y,z\}} S_{tr}^{\gamma\gamma}(\mathbf{k},\omega)$ at low energies for the WCC phase by ED on a 24-site cluster ($3\times 4$ unit cells). In order to capture the dynamics we exclude the ground state contribution.  The results at the $\omega = 0.018$ cut for $K_z/K = 1,~h/K = 1.0 < h_{c2}^{\rm ED}/K \approx 1.1$ are shown in Fig. \ref{fig:dyn}, where one-dimensional patterns are obvious in both $S(\mathbf{k},\omega)$ and $S_{tr}(\mathbf{k},\omega)$. 
For $S(\mathbf{k},\omega)$ the intensity peaks are parallel to $\hat{k}_y$, hence the dynamics is strongly one-dimensional along zigzag compass chains despite strong spin exchange on z bonds. Further, the intensity is primarily concentrated about the second Brillouin zone which indicates the flux (gauge) degrees of freedom remain deconfined and is responsible for the short range correlations between spins \cite{Baskaran2007,Feng2022pra}. 
$S_{tr}(\mathbf{k},\omega)$ in Fig. \ref{fig:dyn}(d) further reveals the emergent 1D nature of the WCC phase; the intensity modulation is primarily along $k_x$ at low energy, and would only develop minor modulation along $k_y$ in higher energy sector reflecting the weak coupling between zigzag chains, which is expected to vanish as $h \rightarrow h_{c2}$ 
Note that the intensity of the dynamical structure factor $S_{tr}(\mathbf{k},\omega)$ is the strongest near $k_x = 0 \pmod{2\pi}, ~\forall k_y$. This is consistent with the fact that the zigzag compass chain can be mapped to a critical TFIM with fermion dispersion $\varepsilon(k_x) \propto \sqrt{2 - 2\cos{k_x}}$ \cite{Brzezicki2007}, as well as the mean field solution of majorana dispersion, as will be discussed in the following section (also see Appendix~\ref{sec:MFT}), where the band inside the WCC phase is gapless at the $\Gamma$ point.

\medskip
\section{Parton mean field theory}
\begin{figure*}[t]
    \centering
    \includegraphics[width=\linewidth]{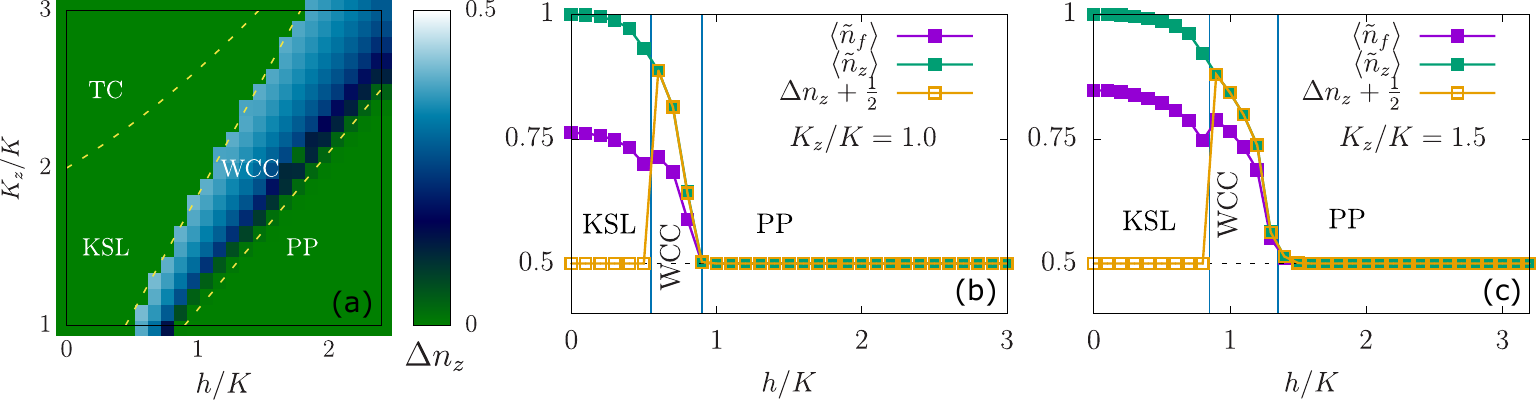}
     \caption{Results from mean field analysis (see Appendix~\ref{sec:MFT}) where the self-consistency relations are solved over a $40\times 40$ momentum grid over the Brillouin zone. (a) Contour plot of $\Delta n_z$ (the difference between $z$ bond occupancy between two consecutive MFT solutions) shows a bistable region reflecting the WCC phase. (b,c) The behavior of $\langle \tilde{n}_f \rangle$ and $\langle \tilde{n}_z \rangle$ and $\Delta n_z$ as a function of $h$ for $K_z/K=1$ and $K_z/K=1.5$ in Kitaev Spin Liquid (KSL), weakly coupled chains (WCC) and partially polarized (PP) phase.}
    \label{phasediagram}
\end{figure*}
We now apply MFT for the parton decomposition applicable for $Z_2$ topological order. Details of the MFT can be found in Appendix C.  Different MFTs for the isotropic Kitaev model under a [001] field have been studied previously~\cite{Liang2018,Nasu2018}. In this section we aim to extend the analysis to the entire $(K_z,h)$ plane and understand the origin of the effectively weakly coupled fermion chains in the WCC phase observed by ED and DMRG as described above.
The spin operator in Eq. \ref{eq_spinHam} is fractionalized into static and itinerant majoranas, denoted by $b$ and $c$ operators respectively, according to the transformation $\sigma^\alpha_{i,A} = i b^\alpha_{i,A} c_{i,A}$. 
The generic Kitaev Hamiltonian under the parton representation is $H_K = \sum_{i, \alpha} K_{\alpha} \Big(  b^\alpha_{i,A}  b^\alpha_{i,A+\hat{\alpha}} c_{i,A} c_{i,A+\hat{\alpha}} \Big)$, 
where $b^\alpha_{i}, c_i$ satisfy the majorana algebra where $c^2_i=1$ and $\{c_i, c_j\}=2\delta_{ij}$. We convert the itinerant majorana fermions into canonical complex fermions $c_{i,A +\hat{z}} = i (f_i-f^\dagger_i),~c_{i,A} =  f_i + f^\dagger_i$ such that $c_{i,A}c_{i,A+\hat{z}} = i(2n^f_i-1)$; and similarly the static majoranas into $b^\alpha_{i,A} = \eta_{i\alpha} + \eta^\dagger_{i\alpha},~b^z_{i,A + \hat{\alpha}} = i (\eta_{i\alpha} - \eta^\dagger_{i\alpha})$ such that $b^\alpha_{i,A}b^\alpha_{i,A + \hat{\alpha}}= i(2n^{\alpha}_i-1)$.
Under these transformations the Kitaev exchanges on $z$ and $x$ bonds become
\begin{align}
    \sigma_{i,A}^z \sigma_{i,A+\hat{z}}^z &=  (2n^f_i-1) (1-2n^{z}_i) \label{eq:niz}\\
    \sigma_{i,A}^x \sigma_{i,A+\hat{x}}^x &= (1-2n^{x}_i)(f_i f_{i-\mathbf{n}_1}-f_if^\dagger_{i-\mathbf{n}_1}+ {\rm H.c.}) \label{eq:nif}
\end{align}
whereas the $K_y$ exchange can be obtained simply by exchanging the indices $\hat{x}\leftrightarrow \hat{y},~ \mathbf{n}_1 \leftrightarrow \mathbf{n}_2$; and the external field along $\alpha$ direction is given by $h^\alpha (i b^\alpha_{i,A} c_{i,A}) = i h^\alpha (\eta_{i\alpha}+ \eta^\dagger_{i\alpha})(f_i + f^\dagger_i)$. In the flux-free sector, bond fermions are uniformly occupied by $n_o=n^z_o=1$. 
Since the majorana sector of the Kitaev model has $p$-wave pairing \cite{Read2000}, and the ED results (Fig. \ref{fig:dyn}) indicate a picture of decoupled TFIM chains, we keep in our mean field ansatz the superconducting correlator $\phi = \expval{f_i f_j}$ and the Hartree term $\xi = \expval*{f_i^\dagger f_j}$ their complex conjugates, and bond fermion occupation $n_0^\alpha = \expval{n^\alpha}$ with $\alpha = x,y,z$. 
The parton decomposition we here perform allows tracking fluctuations in gauge and majorana sectors separately; and capturing strong non-local gauge fluctuation which might connect different topological sectors, as is required for determining the dimensional reduction which changes the effective lattice geometry globally.
This differs from the previous investigation \cite{Liang2018} using the Hartree-Fock decomposition which neglected the $p$-wave (spinless) pairing field $\phi$; and differs from the MFT based on Jordan-Wigner particles \cite{Nasu2018} which is not suitable for detecting emergent decoupled chains due to a non-local Jordan-Wigner transformation \footnote{For example, a stack of decoupled 1D Ising chains can become an intrinsically 2D model with topological order after (inverse) Jordan-Wigner-type transformation, as is discussed in \cite{Chen2007}. }.    
The mean field phase diagram under $h$ field is shown in \Fig{phasediagram}. Three distinct regions can be seen -- and the phase transition points approximately agree with those obtained by ED. The regions are identified based on the gap to the fermionic excitations and on the occupancy of the $\eta_z$ fermion. 
The phases can be identified in terms of fermionic phases for $f$ fermions as follows: (i) the KSL phase is identified as a $p+ip$ nodal superconductor, (ii) the WCC phase as a weakly coupled one-dimensional $p$ wave superconductor, (iii) partially polarized magnetic phase with one-dimensional spin wave excitations, and (iv) TC as a gapped superconductor whose $n_z$ occupation is the same as that in KSL.

At large $K_z$ and weak field, the ground state is given by TC, as shown in Fig. \ref{phasediagram}. In the TC phase, the occupation number of $z$ bond fermion $n^z_i \sim 1$, indicative of the $Z_2$ gauge theory with zero flux on each plaquette. To leading order the [001] field does not alter the low energy subspace of TC, therefore the toric code ground state is stable against a Zeeman field perturbation $h\sigma^z$.  It is only at large $h$ that the $z$ polarized state can be stabilized - hence one expects a direct first order transition between the TC ground state (with zero magnetization quantum number) to the completely polarized state at a scale between $h \sim K_z$. This is in sharp contrast to the [111] field which induces a one-dimensional dispersion of abelian anyons at a critical field $h_{[111]} \sim O(K_z^{-1})$ \cite{Feng20222}. 
At lower anisotropy there are two non-trivial phases, the KSL and WCC phase, which we now discuss.  In the KSL phase the hybridization between the $z$ bond fermions and the $f$ fermions leads to an overall renormalization of the band structure, and $n_z \sim 1$ signals a robust KSL. Note that the KSL phase is an effective nodal gapless superconductor since the $h\sigma^z$ field does not open the majorana gap in contrast to $h_{[111]}$ that does open a gap $\Delta \sim h_{[111]}^3$. 

With further increase of $h$ beyond $h_{c1}$ one finds the interesting WCC phase where mean field solutions oscillate between two saddle points: the solutions of $n_z$ fluctuate between two values $0.5 \pm \Delta n_z$ with the variance $\Delta n_z \neq 0$, and they average to $\langle n_z \rangle \sim 0.5$. It is interesting to point out that everywhere else in the phase diagram the mean field converges to a unique solution. 
In WCC, the bistable mean field solutions reflects the strong quantum fluctuation in the gauge sector. This also makes sense if one considers the fact that the KSL has different topological super-selection sectors. Our MFT allows a global change of the gauge field (e.g. flipping all signs of the $Z_2$ gauge along Wilson lines). Hence the unstable solution can be attribute to very strong fluctuation in the gauge sector or the bond fermion or gauge sector, such that different super-selection sectors can fluctuate into each other due to the large fluctuation of the gauge field.

In order to characterize this bistable region we investigate $ \langle \tilde{n}_z\rangle  \equiv |\langle n_z - \frac{1}{2} \rangle|+\frac{1}{2}$, which tracks the higher branch of $n_z$ solution; and $\Delta n_z$ which tracks the difference between two consecutive mean field solutions (see Fig.~\ref{phasediagram}). Regions where $\Delta n_z = 0$ signal a unique saddle point where $\langle \tilde{n}_z\rangle = \langle n_z\rangle$. We find an extended region in the phase diagram where $\Delta n_z \neq 0$ even while $ \langle \tilde{n}_z\rangle$ is finite, and is qualitatively in the same region where the ED/DMRG shows WCC phase (see Fig.~\ref{EDDMRG}). Note that when $\langle n_z\rangle=0.5$, it signals decoupled Ising chains where $(1-2n_i^z) \sim 0$ in Eq.~\ref{eq:niz}. Thus the WCC region is characterized by significant fluctuations reflected in the multiple stabilities ($\Delta n_z \neq 0$), signaling significant flux excitations; and a negligible ensemble average of $(1-2n_i^z)$, indicating the weakly coupled nature of WCC.  That the chains are weakly coupled due to renormalized $z$ bond fermions is consistent with ED results where low energy dynamics is virtually one-dimensional. 
Thus the KSL-WCC transition is a gapless-to-gapless transition  
from a nodal KSL to WCC akin to a Lifshitz transition in the gapless-to-gapless quantum phase transition in the bilinear-biquadratic spin model \cite{Feng2022PRB,Feng2020PRB}. 
With a further increase in $h$, the system transits into a trivial polarized phase with $\langle n_z\rangle = \langle \tilde{n}_z \rangle=0.5$, where the chains get completely decoupled in the parton mean-field picture, signaling the breakdown of the mean-field ansatz which is to be replaced by a spin wave analysis.
In fact ED results near $h_{c2}$ directly reflects this physics where effective spin exchange $K_z$ along $z$ bonds vanishes in the mean field picture, and the dynamics is thus dominated by fermions hopping along $x$ and $y$ bonds, which is numerically verified in Fig. \ref{fig3}(c), Fig. \ref{fig:dyn}(c,d) and Fig. \ref{fig:eeall}(a,b) in the Appendix \ref{sec:suppres}.

\begin{figure}[t]
    \centering
    \includegraphics[width=0.48\textwidth]{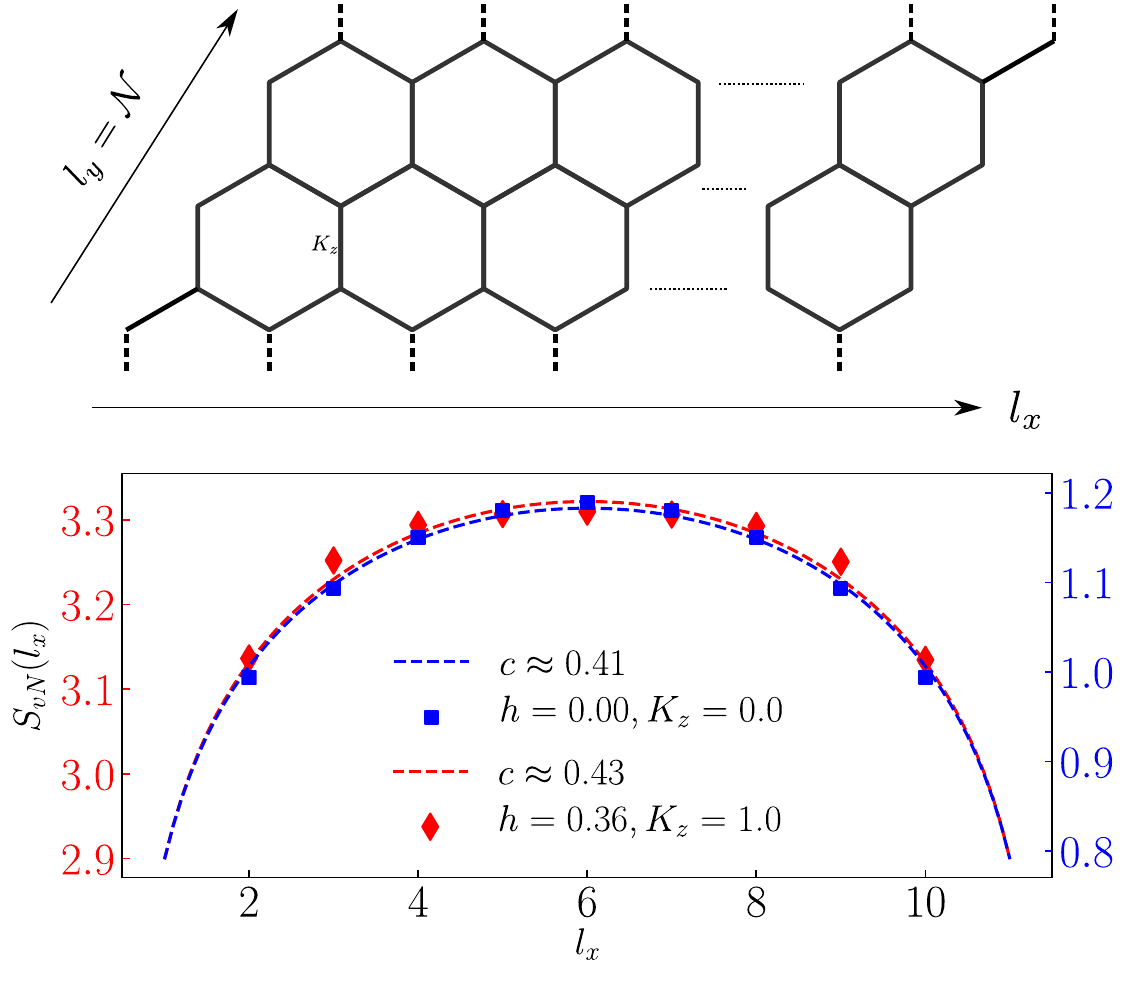}
    \caption{Top: Cylindrical geometry of an $\mathcal{N}$-legged honeycomb ladder. Bottom: scaling of entanglement entropy as a function of $l_x$ at $K_z = 0, h = 0$ (decoupled compass chain), and $K_z = 1, h = 0.71\approx h_{c2}^-$ (emergent 1D Ising criticality inside the intermediate phase). Data obtained by DMRG on a $72$-site, $\mathcal{N} = 3$ ladder ($12\times 3$ unit cells), where $h_{c2}\simeq 0.76$, with maximal bond dimension $4000$ and truncation error $\sim 10^{-8}$.   }
    \label{fig:cc}
\end{figure}

Furthermore, the quasi 1D and critical nature of the intermediate phase close to $h_{c2}$ can be reflected in the scaling of von-Neumann entanglement entropy as a function of subsystem size \cite{HOLZHEY1994443,Vidal2003,Vidal2004,Cardy2004,Calabrese_2009}. For a circular system it is given by $S_{\rm vN}(x) \simeq \left(\frac{c}{3} \right)\log\left[ \frac{x}{a} \right] + \mathcal{C}$, 
with $x$ the length of a segment and $\mathcal{C}$ a constant term. In particular, for a finite size system of an $\mathcal{N}$ legged ladder consisting of $\mathcal{N}$ weakly coupled or decoupled gapless fermion chains, i.e. the subsystem is a union of $\mathcal{N}$ disjoint intervals, the entropy scaling would become
\begin{equation}
S_{\rm vN}(l_x) \simeq \mathcal{N}\frac{c}{6} \log\left[ \frac{2L}{\pi} \sin\left(\frac{\pi l_x}{L}\right) \right] + \mathcal{C} \label{eq:eescaling}
\end{equation}
where $l_x$ in our setup is the length of the cylindrical subsystem along $x$ direction of a three-legged ladder measured from the boundary, and $\mathcal{C}$ a constant that does not depend on subsystem size $l_x$. 
We numerically verified that the intermediate phase has $\mathcal{N}c \simeq \frac{\mathcal{N}}{2}$ (or $c \simeq \frac{1}{2}$ per chain) by DMRG, as is shown in Fig. \ref{fig:cc}. Remarkably, the curvature of $S_{\rm vN}(l_x)$ of the decoupled critical compass chain ($K_z = 0$) is virtually the same as that of the Kitaev model (e.g. $K_z = K = 1$) subjected to $h \simeq h_{c2}^-$ up to a constant shift, supporting their equivalence despite the finite-size-induced deviation of the exact value of the fitted central charge \footnote{Note that the decomposition into separated gauge and Majorana, thus the decoupling of z bonds in the mean field picture, do not hold on the boundaries of the cylindrical ladder. Hence, the boundary spins are still entangled along z bonds despite of the decoupling in the bulk, resulting in a smaller entanglement between the boundary spins and the bulk spins due to the monogamy nature of entanglement. Therefore, in order to catch the decoupled nature of the bulk spin chains in a finite size DMRG we need to fit the $S_{\rm vN}(x)$ while avoiding the boundary spins.}. 

\section{Discussion and conclusion}
We have shown that WCC phase is dominated by fermions propagating along the zigzag direction, it is then left to ascertain the universality class of the 1D gapless theory, and discuss the extent to which the effective 1D theory remains valid. If we neglect the $K_z$ coupling inside the WCC phase, the model becomes a 1D compass chain with alternating $K_x$ and $K_y$ spin exchange interactions. The 1D compass chain can then be mapped to the critical point of the TFIM, whose dynamics is governed by gapless free fermions \cite{Brzezicki2007,Olaussen2005}. 
The gapless phase is therefore governed by a (1+1)D CFT with central charge $c=(\frac{1}{2},\frac{1}{2})$ relevant for the left and right moving majorana particles: $\gamma_L$ and $\gamma_R$. The linear energy density of WCC, i.e. the Hamiltonian of one of the zigzag chains, is then dominated by the chiral majorana fields $\mathcal{H}_{L} = \int dx (i\gamma_L \partial_x \gamma_L ),~ \mathcal{H}_{R} = \int dx (-i\gamma_R \partial_x \gamma_R )$. 
In the presence of a weak interchain coupling, which can be approximated to the leading order by the coupling between the two majorana modes, the effective Hamiltonian is given by 
\begin{equation}
    \mathcal{H}({\rm WCC}) \approx \mathcal{H}_{L} + \mathcal{H}_{R} - g\int dx (\gamma_L \partial_x \gamma_L)(\gamma_R \partial_x \gamma_R) \label{eq:eff}
\end{equation}
where $g$ is a parameter that determines the leading self-energy of $\gamma$ due to the existence of interchain coupling. The effect of the interchain coupling $g$ is to gap out the bulk majorana modes leading to the emergence of a chiral Ising edge CFT which is a universal property of the non-abelian spin liquid \cite{Liu2022}. It is known that the $c = \frac{1}{2}$ phase of Eq. \ref{eq:eff} remains robust against perturbations in $g$ up to a finite critical $g_c$ \cite{Affleck2015a,Affleck2015b,Rahmani_2019}. Therefore, we expect that the WCC phase remains an effective model of 1D fermion chains under the presence of weak interchain coupling, as is described in MFT and also supported by our numerical calculations. Since the coupling between these fermionic chains is weak, it is expected that for low temperatures the system's behavior would be describable by the quantum Ising criticality ~\cite{Qimiao2018}. Furthermore, our work demonstrates the tunability by magnetic field between the $Z_2$ Ising critical point and the Kitaev QSL. There have been several proposals whereby QSLs can be boostrapped from $Z_2$ Ising critial chains by turning on the inter-chain coupling between left and right moving (majorana) fermions \cite{Mudry2017,Slagle2022,liu2023}, which can be viewed as a reverse process of our finding where Kitaev QSL is reduced to decoupled $Z_2$ critical fermion chains. 

In summary, we have presented the unusual physics of dimensional reduction generated by a [001] uniform magnetic field in an otherwise two-dimensional Kitaev spin liquid. The phase diagram of the anisotropic Kitaev model under a [001] field is in sharp contrast to that induced by a [111] field. A weak [001] field induces the gapless KSL phase to remain a gapless nodal superconductor, which transits to an intermediate gapless phase dubbed WCC (weakly coupled chains) at $h_{c1}$. Remarkably this phase persists even for relatively strong anisotropy under a finite field $h \lesssim 1.1K_z$. We showed by MFT paradigm that the transition is described as a Lifshitz-type transition driven by the reduction of $z$-type bond fermions. The intermediate gapless phase induced by the [001] field can be captured by an effective model of weakly coupled 1D compass chains, which asymptotically approaches a (1+1)D CFT with $c = \frac{1}{2}$ at $h_{c2}$, as is supported by various numerical calculations near $h_{c2}$. Finally, in the trivial PP phase $h > h_{c2}$, the model becomes completely decoupled bosonic chains under the LSW theory. Due to the three-fold spatial rotational symmetry ($\theta = 2\pi/3$) of the original Kitaev Hamiltonian, such argument holds as well for [010] and [100] fields. 

With the advent of recent experiments shedding light on field-induced quantum criticality under the influence of a magnetic field \cite{Wolter2017,Nagai2020}, and theoretical advancements unveiling novel phases in Kitaev systems \cite{Feng20222,zhang2023machine}, the field-induced phenomena in QSLs is an area of active research. 
Particularly intriguing is the recent discovery of a soliton phase in 1D Kitaev spin chains \cite{Kee2023,Kee2023b}, which bears significant relevance to the dimensional reduction from 2D Kitaev systems to 1D Ising chains explored in our work. 
Our work underlines the rich interplay between dimensionality and criticality and we anticipate our findings will guide future experimental investigations of spin liquid phases under external fields.

\medskip
{\bf {Acknowledgement.}} We acknowledge N. D. Patel, S. Bhattacharjee, A. Nanda for discussions and collaborations on related projects. S. Feng acknowledges support from NSF Materials Research Science and Engineering Center (MRSEC) Grant No. DMR-2011876 and the Presidential Fellowship of The Ohio State University; N. Trivedi acknowledges support from NSF-DMR 2138905; A. Agarwala acknowledges support from IIT Kanpur Initiation Grant IITK/PHY/2022010. S. Feng expresses gratitude to the Boulder Summer School 2023 at the University of Colorado Boulder for useful lectures and discussions related to topics in this project.

\appendix

\section{Linear spin wave theory} \label{sec:lsw}
Here we approach the WCC phase from the high field polarized limit. 
In PP phase induced by [001] field, LSW disperses only along the zigzag direction, while exhibits a flat band along the armchair direction due to the absence of boson hopping terms on $z$ bonds, which is in sharp contrast to topological magnons induced by [111] field or Heisenberg exchange \cite{Joshi2018}, and was previously alluded to in the context of the square-octagon Kitaev lattice \cite{Hickey20212}. Such virtual dimensional reduction is visualized in Fig. \ref{fig3}(e,f), where it is readily seen from the suppression of propagation along the armchair direction. This can be made explicit by the effective LSW Hamiltonian in the PP. We apply the Holstein–Primakoff which maps spin-1/2 operators to bosons as fluctuations from the ordered magnetic moments. For low magnon density $\expval*{a_i^\dagger a_i}/2S \ll 1$ we can keep only upto the first order as follows: $S_i^z = S - a_i^\dagger a_i$, $S_i^+ \approx \sqrt{2S}(1 - \frac{a_i^\dagger a_i}{4S}) a_i$ and $S_i^-\approx \sqrt{2S}a_i^\dagger (1 - \frac{a_i^\dagger a_i}{4S})$, 
so that to the leading order which contributes to boson bilinears, $S_i^x$ and $S_i^y$ becomes $S_i^x \approx \sqrt{\frac{S}{2}}(a_i + a_i^\dagger),~ S_i^y \approx -i\sqrt{\frac{S}{2}}(a_i - a_i^\dagger)$.
Let $a$ and  $b$ denote the boson operator for $A$ and $B$ sublattice respectively, we now enumerate all couplings in Kitaev honeycomb Hamiltonian upto quadratic order:
\begin{align}
	S_{i,A}^x S_{j_1,B}^x &= \frac{S}{2}(a_i b_{j_1} + a_i^\dagger b_{j_1}^\dagger + a_i b_{j_1}^\dagger + a_i^\dagger b_{j_1})\\
	S_{i,A}^y S_{j_2,B}^y &= \frac{S}{2}(a_ib_{j_2}^\dagger + a_i^\dagger b_{j_2} -a_i b_{j_2} - a_i^\dagger b_{j_2}^\dagger)\\
	S_{i,A}^z S_{i,B}^z &= S^2 - S a_i^\dagger a_i - S b_{i}^\dagger b_{i} \label{eq:zz}
\end{align}
where $j_1 = i+\mathbf{n}_1,~ j_2 = i+\mathbf{n}_2$. The magnetic field in $z$ diection is
\begin{equation} \label{eq:zfield}
	-h (S_{i,A}^z + S_{i,B}^z) = -h (2S - a_i^\dagger a_i - b_{i}^\dagger b_{i})
\end{equation}
From Eq. \ref{eq:zz} and Eq. \ref{eq:zfield} it is already clear that bosons do not hop along $z$ bonds, resulting in the flat band along $k_y$ cuts in the Brillouin zone. 
We apply the Fourier transformation
\begin{equation}
	a_j = \frac{1}{\sqrt{N}} \sum_{\mathbf{k}} e^{i\mathbf{k}\cdot \mathbf{r}_j} a_{\mathbf{k}},~~ b_l = \frac{1}{\sqrt{N}} \sum_{\mathbf{k}} e^ {i\mathbf{k}\cdot \mathbf{r}_l} b_{\mathbf{k}} 
\end{equation}
where we assume $2N$ spins. It is then readily to get the LSW Hamiltonian. By plugging in $S=1/2$ and dropping constant we have
\begin{equation}
	\begin{split}
		H = &\sum_{\mathbf{k}}  \frac{1}{4}\left( K_x e^{i\mathbf{k}\cdot \mathbf{n}_1} - K_y e^{i\mathbf{k}\cdot \mathbf{n}_2} \right)(a_\mathbf{k} b_{-\mathbf{k}} + a_\mathbf{k}^\dagger b_{-\mathbf{k}}^\dagger) \\ 
			      &+ \frac{1}{4} \left( K_x e^{i\mathbf{k}\cdot \mathbf{n}_1} + K_y e^{i\mathbf{k}\cdot \mathbf{n}_2}\right) (a_\mathbf{k} b_\mathbf{k}^\dagger + a_\mathbf{k}^\dagger b_\mathbf{k}) \\
			      &+ \frac{1}{2}\left(2h - K_z\right)(a_\mathbf{k}^\dagger a_\mathbf{k} + b_\mathbf{k}^\dagger b_\mathbf{k}) 
	\end{split}
\end{equation}
which in the block form after symmetrization is written as
\begin{equation} \label{eq:lsw}
	H_{\rm LSW} = \frac{1}{2}\sum_{\mathbf{k}} \Psi_\mathbf{k}^\dagger \mathbf{H}(\mathbf{k}) \Psi_\mathbf{k} 
	,~
	\mathbf{H}(\mathbf{k}) \equiv 
	\begin{pmatrix}
		\mathbf{M}(\mathbf{k}) & \mathbf{N}(\mathbf{k}) \\
		\mathbf{N}^\dagger(\mathbf{k}) & \mathbf{M}(-\mathbf{k}) \\
	\end{pmatrix}
\end{equation}
where we defined $\Psi_\mathbf{k} \equiv (a_\mathbf{k}, b_\mathbf{k}, a_{-\mathbf{k}}^\dagger, b_{-\mathbf{k}}^\dagger)^{\rm T}$ with $a,b$ boson operators of $A$ and $B$ sublattices, and the two-by-two matrices $\mathbf{N}(\mathbf{k})$ and $\mathbf{M}(\mathbf{k})$ are defined by
\begin{align}
	    \mathbf{N} &= \frac{1}{4} \left( K_x e^{i\mathbf{k}\cdot \mathbf{n}_1} - K_y e^{i\mathbf{k}\cdot \mathbf{n}_2} \right) \sigma^x\\
    	\mathbf{M} &= (h - \frac{1}{2}K_z)\sigma^z + \frac{1}{4} \left( K_x e^{i\mathbf{k}\cdot \mathbf{n}_1} + K_y e^{i\mathbf{k}\cdot \mathbf{n}_2} \right)\sigma^x \label{eq:M}
\end{align}
which, in the real space, do not contain boson hopping terms along $z$ bonds. This is apparent from the fact that $K_z$ is associated only with $\sigma^z$ in Eq. \ref{eq:M}, which is not coupled to momentum. The resulting dispersion of LSW in shown in Fig. \ref{fig:dyn}(b), which is qualitatively consistent with ED result in Fig. \ref{fig3}(e). It is in sharp contrast to the anti-ferromagnetic Heisenberg model on a honeycomb lattice where no dimensional reduction is present under arbitrary Zeeman field \cite{Fransson2016}. 
Furthermore, the LSW theory also provides an estimation of the critical field $h_{c2}$ in the thermodynamic limit: The energy contribution of boson occupation number on z bonds changes sign when the first term in Eq. \ref{eq:M} changes sign. 
Hence from Eq. \ref{eq:M} we see the singularity between WCC and PP phase occurs at $h_{c2} \sim K_z$, which is very close to the numerical result shown in Fig. \ref{fig:dyn}(a) (Note there is a difference in $h/K$ by a factor of two between spin-$\frac{1}{2}$ and Pauli operator).
Therefore, the boson energy density within z bonds asymptotically vanishes as the magnetic field approaches the critical point $h\rightarrow h_{c_2} \sim K_z$ from within PP phase, and the effective model transition from weakly coupled fermionic chains to that of virtually decoupled bosonic chains. This is also consistent with $h \rightarrow h_{c2}$ from within the WCC phase, where the $z$-bond coupling becomes negligible in the WCC phase

\begin{figure}[t]
    \centering
    \includegraphics[width=\linewidth]{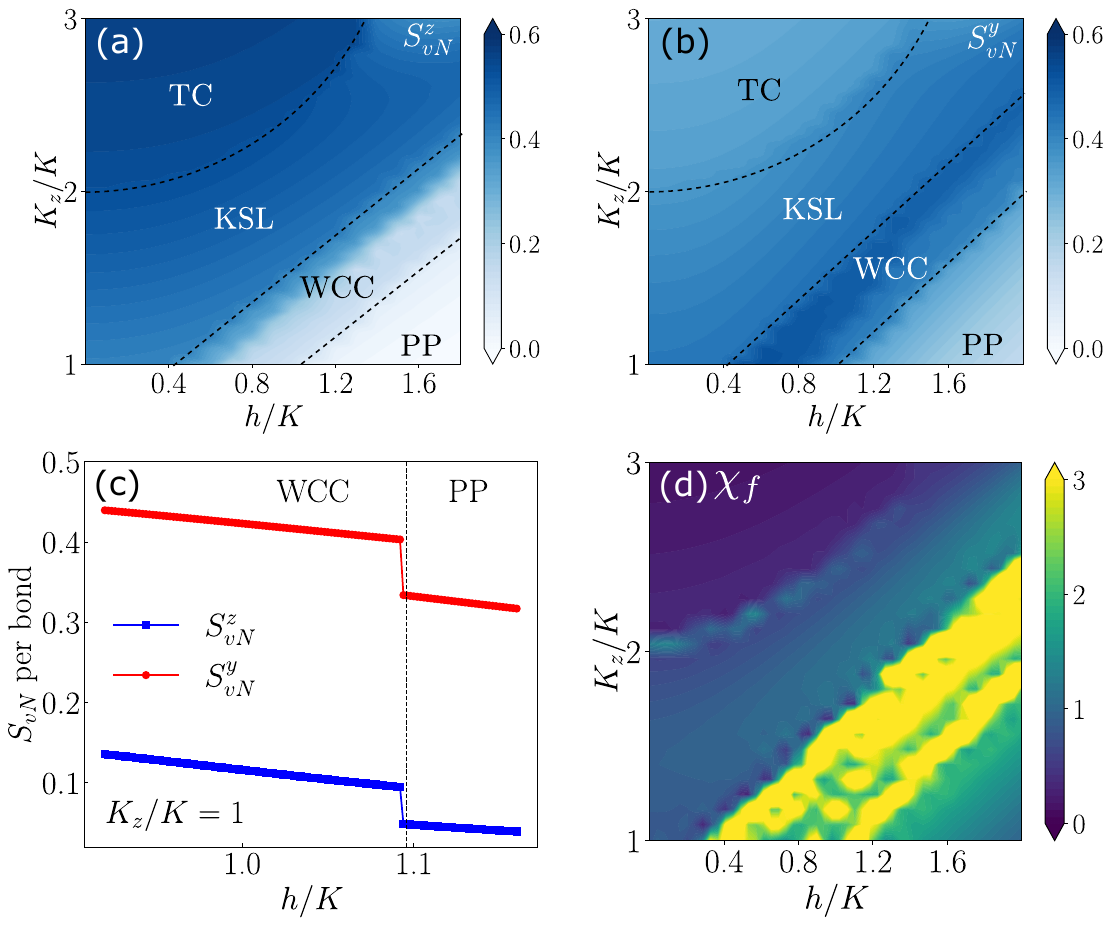}
    \caption{Phase diagram measured by entanglement entropy per bond and susceptibilities.  (a) $S_{\rm vN}^z$ -- entanglement entropy of subsystem whose boundary cuts through only $z$ bonds; and (b) $S_{\rm vN}^y$ -- entanglement entropy of subsystems whose boundary cuts througth only $y$ bonds. (c) The zoomed-in profile of $S_{\rm vN}^z$ and $S_{\rm vN}^y$ near $h_{c2}^{\rm ED}$ with $K_z/K = 1$; the vertical dashed line marks out $h_{c2}^{\rm ED}/K \approx 1.1$. (d) Fidelity susceptibility. Data obtained by 24-site ED with torus geometry. }
    \label{fig:eeall}
\end{figure}
\section{Additional entanglement results}
\label{sec:suppres}
In this appendix we present additional numerical results. We first discuss the entanglement entropy of the phase diagram as a function of $(K_z/K,h/K)$. In order to compare the entanglement of different cuts, i.e. how entangled the ground state is along different directions, we reduce that dimension and calculate the von-Neumann entanglement entropy per bond. The von-Neumann entanglement entropy per bond of the two cuts used in the main text is given by
\begin{align}
    S_{\rm vN}^i &= -\frac{1}{|\partial A_i|}\Tr\rho_{A_i} \ln \rho_{A_i},~ i \in \{z,y\}
\end{align}
where $\rho_{A_z}$ and $\rho_{A_y}$ are reduced density matrices of subsystems whose boundaries cut through only $z$ bonds and $y$ bonds respectively, and $|\partial A_i|$ denotes the length of the boundary of the subsystem. The calculated entropy per $z$ or $y$ bond is shown in Fig. \ref{fig:eeall}(a,b). The zoomed-in profile of the entanglement entropy is shown in Fig. \ref{fig:eeall}(c), which marks out the $h_{c2}^{\rm ED}/K \approx 1.1$. 

In the TC phase, $S_{\rm vN}^z$ is significantly larger than $S_{\rm vN}^y$. This is due to that fact that $z$ bonds in TC are much stronger than $x$ and $y$ bonds. Since the majorana gap in TC is a lot larger than the flux gap, the $S_{\rm vN}^z$ of TC can be readily estimated to be $S_{\rm vN}^z\rightarrow\ln2$, and $S_{\rm vN}^y \rightarrow \frac{\ln2}{2}$ when $K_z/K \rightarrow \infty$ \cite{Feng2022pra}. Similarly, in the KSL phase when $h \rightarrow 0$, due to the short range correlation of spins, we can also estimate $S_{\rm vN}^z \sim 0.5$ as is mentioned in the main text. In Fig. \ref{fig:eeall}(a) it is also readily to see that WCC phase features much smaller $S_{\rm vN}^z$ than TC and KSL phases, which, as is discussed in the main text, reflects the weak coupling between zigzag chains. In contrast, Fig. \ref{fig:eeall}(b) shows that $S_{\rm vN}^y$ of WCC is strongly entangled along the zig-zag direction. These results suggest that the WCC phase is a lot more coherent along zigzag chains, yet virtually decoupled between these chains.  

In addition to bipartite entanglement entropy, in Fig. \ref{fig:eeall}(d), we also show the fidelity susceptibility defined by $\chi_f = \partial^2 E_{\rm g.s.}/\partial K_z^2$, which reveals a phase diagram consistent with the that given by the magnetic susceptibility $\chi$ shown in the main text, and that by the entanglement entropy in Fig. \ref{fig:eeall}(a,b). The TC-to-KSL phase transition is made more clear by the measure of $\chi_f$ [Fig. \ref{fig:eeall}(d)] than that by $\chi$ [Fig. \ref{EDDMRG}(c)], since this transition is essentially driven by anisotropy instead of magnetic field. Notably, the WCC phase has significantly larger $\chi$ and $\chi_f$ than all other surronding phases, indicating that WCC is a gapless phase with long-range correlation. This is consistent with that our proposal that WCC is effectively an 1+1D CFT with $c=(\frac{1}{2},\frac{1}{2})$, originating from the 1D compass model which can be mapped to the critical point of TFIM when the interchain coupling is weak.

\begin{widetext}
\section{Parton mean field theory}
\label{sec:MFT}
The spin Hamiltonian with just Kitaev exchanges are given by $H = \sum_{j, \alpha} K_{\alpha} \Big(  \sigma^\alpha_{j,A}  \sigma^\alpha_{j,A+\hat{\alpha}}  \Big)$. Note for convenience we formalize the MFT in Pauli matrices to avoid $\frac{1}{2}$ factors. 
Under the transformation $\sigma^\alpha_{j,A} = i b^\alpha_{j,A} c_{j,A}$ and the static flux sector the Hamiltonian becomes $H = \sum_{j, \alpha} K_\alpha \Big( i c_{j,A} c_{j,A+\hat{\alpha}}  \Big)$. 
We now wish to include the effect of flux excitations as well into the formalism at the mean-field level.
The complete Hamiltonian of our interest is thus 
\beq
H = \sum_{j, \alpha} K_{\alpha} \Big(  \sigma^\alpha_{j,A}  \sigma^\alpha_{j,A+\hat{\alpha}}  \Big) + \sum_{j,\alpha} h (\sigma^{\alpha}_{j,A} + \sigma^{\alpha}_{j,A+\hat{z}})
\eeq 
Using $\sigma^\alpha_{j,A} = i b^\alpha_{j,A} c_{j,A}$ we write the Hamiltonian in terms of these majoranas:
\beq
\begin{split}
    H = \sum_{j, \alpha} K_{\alpha} \Big(  b^\alpha_{j,A}  b^\alpha_{j,A+\hat{\alpha}} c_{j,A} c_{j,A+\hat{\alpha}} \Big) + \sum_j h( i b^\alpha_{j,A} c_{j,A}+ i b^\alpha_{j,A+\hat{z}} c_{j,A+\hat{z}})
\end{split}
\eeq
where $b^\alpha_{j}, c_j$ satisfy the Majorana algebra where $c^2_j=1$ \cite{kitaev2006anyons} and therefore $\{c_j, c_k\}=2\delta_{jk}$. 
Converting these Majorana fermions into complex fermions $c_{j,A} \equiv  f_j + f^\dagger_j,~
c_{j,A +\hat{z}} =i (f_j-f^\dagger_j)$, 
where $f$ satisfy the canonical fermionic algebra. Using this
\begin{align}
    c_{j,A}c_{j,A+\hat{z}}&=i(2n^f_j-1) \\
    c_{j,A}c_{j,A+ \hat{x}} &= i(f_j f_{j-\mathbf{n}_1} + f^\dagger_j f_{j-\mathbf{n}_1} + {\rm H.c.})\\
    c_{j,A}c_{j,A+ \hat{y}} &=i(f_j f_{j-\mathbf{n}_2} + f^\dagger_j f_{j-\mathbf{n}_2} + {\rm H.c. })
\end{align}
where we used the fact that $    c_{j,A}c_{j,A+ \hat{x}} = c_{j,A}c_{j-\mathbf{n}_1,A+ \hat{z}},~
c_{j,A}c_{j,A+ \hat{y}} = c_{j,A}c_{j-\mathbf{n}_2,A+ \hat{z}}$. 
Similarly we have $b^\alpha_{j,A} = \eta_{j\alpha} + \eta^\dagger_{j\alpha}$ and $b^z_{j,A + \hat{\alpha}} = i (\eta_{j\alpha} - \eta^\dagger_{j\alpha})$, 
such that $b^\alpha_{j,A}b^\alpha_{j,A + \hat{\alpha}}= i(2n^{\alpha}_j-1)$.
Under these transformations the Hamiltonian is given by
\begin{align}
    K_z \Big(  b^z_{j,A}  &b^z_{j,A+\hat{z}} c_{j,A} c_{j,A+\hat{z}} \Big) =  K_z (2n^f_j-1) (1-2n^{z}_j)\\
    K_x \Big(  b^x_{j,A}  &b^x_{j,A+\hat{x}} c_{j,A} c_{j,A+\hat{x}} \Big) = K_x (1-2n^{x}_j) 
    (f_j f_{j-\mathbf{n}_1} + f^\dagger_j f_{j-\mathbf{n}_1} + {\rm H.c.})\\
    K_y \Big(  b^y_{j,A}  &b^y_{j,A+\hat{y}} c_{j,A} c_{j,A+\hat{y}} \Big) = K_y (1-2n^{y}_j)
   (f_j f_{j-\mathbf{n}_2} + f^\dagger_j f_{j-\mathbf{n}_2} + {\rm H.c. })
\end{align}
\begin{figure*}[t]
    \centering
    \includegraphics[width=\linewidth]{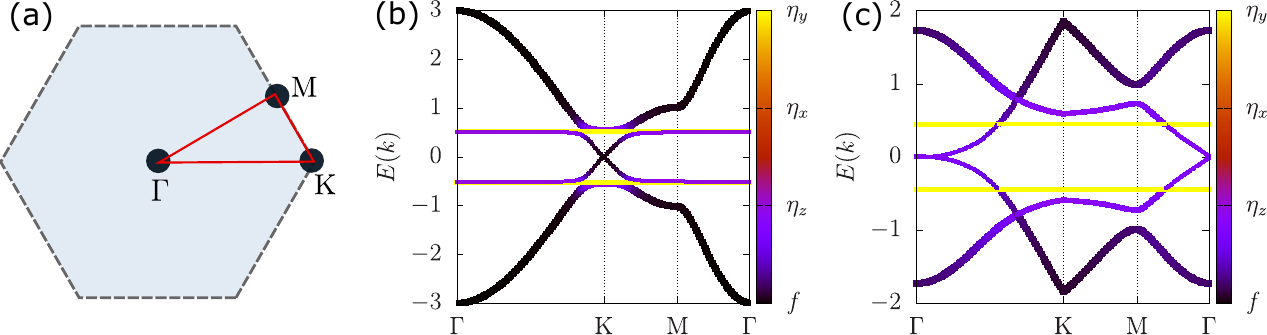}
     \caption{Mean field analysis of dispersion of different partons. (a) The cut in the first Brillouin zone. (b) Dispersion of different particles in the KSL phase ($K_z/K=1, h/K=0.1)$. $f$ fermion is gapless at $\mathbf{K}$ point while bond fermions $\eta_\alpha$s remain gapped. (c) Dispersion of different particles in the WCC phase $(K_z/K=1, h/K=0.7)$. $f$ fermions hybridize with $\eta_z$ and is gapless at $\Gamma$ point. The color bar shows the orbital contribution of the states.}
    \label{disperse}
\end{figure*}
Interestingly at this level we see the system mimics that of four fermionic degrees of freedom interacting on a triangular lattice such that the sign structure of various hopping terms are dependent on the occupancies of $n^{\alpha}$ fermions.
The hybridization between $f$ and $\chi$ fermions is induced by a magnetic field, in this case, along [001] direction. The field dependent terms, for each sublattice in a unit cell, are given by
\bea \label{eq:hp}
h (i b^z_{j,A} c_{j,A}) &= i h (\eta_{jz}+ \eta^\dagger_{jz})(f_j + f^\dagger_j),~~
h(i b^z_{j,A+\hat{z}} c_{j,A+\hat{z}}) &= - ih (\eta_{jz}- \eta^\dagger_{jz})(f_j - f^\dagger_j)
\eea
Define the mean fields: $\expval*{2n_j^f - 1} \equiv n_0^f$, $\expval*{f_j f_{j-\mathbf{n}_1}} = \expval*{f_j f_{j-\mathbf{n}_2}} \equiv \phi$, $n_0^z \equiv \expval{n_j^z}$, $n_0 \equiv \expval{n_j^x} = \expval{n_j^y}$, and $\expval*{f_{j-\mathbf{n}_1}^\dagger f_j} \equiv \xi$, 
and use the mirror symmetry $(\mathbf{n}_1 \leftrightarrow \mathbf{n}_2)$, the total MF Hamiltonian is then given by
\begin{equation}
    \begin{split}
        H^{\rm MF} &= -K_z \left[n_0^f ( 2n_j^z - 1 ) + (2n_j^f - 1) (2n_0^z - 1) \right] - K (\phi + \phi^{*} + \xi + \xi^{*}) [ 2( n_j^x + n_j^y) - 2] \\
        & - K (2n_0 - 1) \left[(f_j f_{j-\mathbf{n}_1}-f_j f^\dagger_{j-\mathbf{n}_1}+f^\dagger_j f_{j-\mathbf{n}_1}-f^\dagger_j f^\dagger_{j-\mathbf{n}_1}) + (\mathbf{n}_1 \leftrightarrow \mathbf{n}_2) \right] + 2ih(\eta_{jz}^\dagger f_j - f_j^\dagger \eta_{jz})
    \end{split}
\end{equation}
Note that since the majorana sector of Kitaev model has $p$-wave pairing, and the ED results indicate a picture of decoupled $p$-wave superconducting chains, we have kept the superconducting correlators. We neglected the Fock term in the mean fields of $H_K$ since it only amounts to renormalizing the magnetic field (Eq.~\ref{eq:hp}) by a constant.  
Moving to momentum space by $f_j = \frac{1}{\sqrt{N}}\sum_\mathbf{k} e^{i\mathbf{k}\cdot \mathbf{r}_j} f_{\mathbf{k}}$ and $\eta_{j\alpha} = \frac{1}{\sqrt{N}}\sum_\mathbf{k} e^{i\mathbf{k}\cdot \mathbf{r}_j} \eta_{\mathbf{k}\alpha}$, we have
\begin{equation}
\begin{split}
        H^{\rm MF}(\mathbf{k}) &= - K_z \left[ n_0^f (2 n_\mathbf{k}^z - 1) + (2 n_\mathbf{k}^f - 1)(2 n_0^z - 1)\right] - K (\phi + \phi^{*} + \xi + \xi^{*}) [ 2( n_\mathbf{k}^x + n_\mathbf{k}^y) - 2]  \\
        & -K(2n_0 - 1)\sum_{l=1,2} \left[ \cos(\mathbf{k}\cdot \mathbf{n}_l) n_\mathbf{k}^f + i\sin(\mathbf{k}\cdot\mathbf{n}_l) f_\mathbf{k} f_{-\mathbf{k}}+ h.c.\right] + 2ih (\eta_{\mathbf{k}z}^\dagger f_\mathbf{k} - f_\mathbf{k}^\dagger \eta_{\mathbf{k}z})
\end{split}
\end{equation}
In particular, at $h=0$, where the low energy sector is given by setting $n_o=n^z_o=1$. At this level $f$ and $\eta$ live in separate quantum sectors, and the MFT is exact. 
Hence the four matrix elements of the two-by-two MFT Hamiltonian of $f$ fermions is given by
\begin{align}
    &\makebox[0pt]{$H_{11}^{\rm MF} = \frac{1}{2}\Big(-2K_z(2n_0^z - 1)  - 2K(n_0 - 1) \left[ \cos (\mathbf{k}\cdot \mathbf{n}_1 ) + \cos(\mathbf{k}\cdot \mathbf{n}_2)\right]  \Big) f^\dagger_k f_k$} \\
    &\makebox[0pt]{$H_{22}^{\rm MF} = -\frac{1}{2}\Big(-2K_z (2n_0^z - 1) - 2K(n_0 - 1) \left[ \cos (\mathbf{k}\cdot \mathbf{n}_1 ) + \cos(\mathbf{k}\cdot \mathbf{n}_2)\right]  \Big)  f_{-k} f^\dagger_{-k}$} \\
    &\makebox[0pt]{$H_{12}^{\rm MF} = -i K (n_0 - 1) \left[\sin(\mathbf{k}\cdot \mathbf{n}_1) + \sin(\mathbf{k}\cdot \mathbf{n}_2)  \right] f^\dagger_k f^\dagger_{-k}$}\\ 
    &\makebox[0pt]{$H_{21}^{\rm MF} = iK (n_0 - 1) \left[\sin(\mathbf{k}\cdot \mathbf{n}_1) + \sin(\mathbf{k}\cdot \mathbf{n}_2)  \right] f_{-k} f_{k}$}
\end{align}
\end{widetext}
where $\mathbf{n}_1 = (\frac{1}{2} , \frac{\sqrt{3}}{2})$, $\mathbf{n}_2 = (-\frac{1}{2}, \frac{\sqrt{3}}{2})$, $K = K_x = K_y$. 
At $h=0$, as $K_z$ is increased the $f$ fermions undergo a transition at the $M=\{0,\frac{2\pi}{3}\}$ at $K_z=2K$, where the dispersion looks semi-Dirac like \cite{Burnell2011,Feng20222}, given that the gauge configuration is chosen as $n^\alpha_i = 1$. In presence of magnetic field, the chemical potential of these fermions are effectively renormalized, changing the band structure and the location of gapless modes.  Parton dispersion obtained by such MFT is shown in Fig.~\ref{disperse}, where we take the momentum cut along $\Gamma - \rm K - \rm M - \Gamma$ as presented in Fig. \ref{disperse}(a).  As is shown in Fig. \ref{disperse}(b,c), where, the (hybridized) gapless fermion mode has moved from the $\pm \mathbf{K}$ points in the KSL phase [Fig. \ref{disperse}(b)] to the $\Gamma$ points in the WCC phase [Fig. \ref{disperse}(c)], which is consistent with the ED results where the dynamical structure factor has its strongest weight at $k_x = 0$.

\section{Finite size analysis} \label{app:finite}
\begin{figure*}
    \centering
    \includegraphics[width=\linewidth]{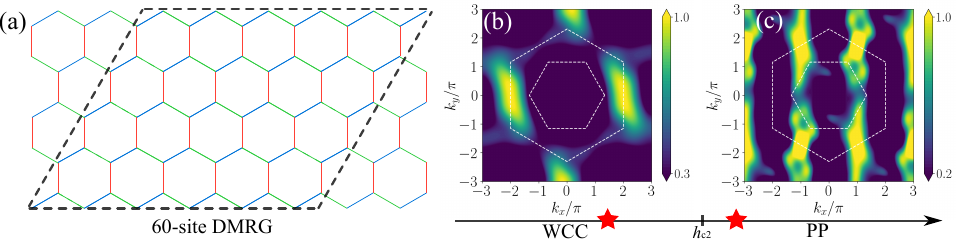}
    \caption{(a) The cylinder geometry used for the 60-site DMRG ($6\times 5$ unit cells), where $h_{c2}\simeq 0.68$, with maximal bond dimension $1200$ and truncation error $\sim 10^{-8}$. (b) The interpolated ground-state static structure factor $S(\mathbf{k})$ of WCC phase at $(K_z/K = 1, h/K = 0.50)$, as defined in Eq. \ref{eq:ss}.  White dashed lines mark the first and extended Brillouin zones. (c) the same quantity in PP at $(K_z/K = 1, h/K = 0.72)$. The agreement between the smaller-scale ED calculations in Fig.~\ref{EDDMRG} under PBC and the 60-site DMRG under cylindrical geometry lends credibility to the persistence of this emergent one-dimensional physics. }
    \label{fig:appD}
\end{figure*}
Now we introduce some additional data that mitigate fintie size effects. Based on further DRMG results, we remark that both the structure factor and the central charge tend to converge on the conclusion that the emergent one-dimensional physics persist in large system sizes. 

{\it Structure factor:}  
In the main text, we presented the static structure factor as calculated using ED on a 24-site system with PBC. Figure~\ref{EDDMRG}(b) highlights that within the WCC phase, the dominant signal is distinctly one-dimensional, aligning with the zigzag direction; and a different one-dimensional signature present in the polarized phase, suggesting magnons with constraint mobility along the same zigzag direction under the high field limit. An open question is whether this one-dimensional characteristic persists in larger systems. To address this, we employed DMRG on a larger cylinder geometry — OBC along the x-direction and PBC along the y-direction, as depicted in Fig.~\ref{fig:appD}(a). Our findings reveal that the intermediate WCC phase maintains effective one-dimensional fluctuations even in a 60-site cylindrical system, as is shown in Fig.~\ref{fig:appD}(b): at intermediate field $h=0.50$ ($h_{c1} < h = 0.50 < h_{c2} \approx 0.68$), the signal aligns with the vertical boundaries of the second Brillouin zone. 
Furthermore, Fig.~\ref{fig:appD}(c) shows that at larger field, the scattering signal of the PP phase also becomes effectively one-dimensional, suggesting the constraint mobility of magnons. These observations align with the static structure factor obtained from the 24-site ED with PBC, illustrated in Fig.~\ref{fig3}(c,e); and are consistent with low energy dynamics of WCC and PP phases, as shown in Fig.~\ref{fig:dyn}(b,c) of the main text. The agreement between the smaller-scale ED calculations under PBC and the larger DMRG computations under cylindrical geometry lends credibility to the persistence of this emergent one-dimensional physics. Although further validation with large-scale DMRG would be ideal, such calculations pose significant challenges in gapless systems. Nonetheless, the observed consistency across disparate geometrical configurations and system sizes strongly supports the robustness of the emergent one-dimensional physics.

{\it Central charge:} Doing DMRG on a gapless system is technically difficult in 2D systems. To the best of our effort, we can reach the stable results presented in Fig.~\ref{fig:cc} with bond dimension 4000 for the 2D cylinder. 
To make sense of the finite size error, we have looked into the fitted central charge in smaller system sizes: $36$-${\rm site~} (L_x = 6) \rightarrow c \approx 0.33$, $54$-${\rm site~} (L_x = 9) \rightarrow c \approx 0.39$, $72$-${\rm site~} (L_x = 12) \rightarrow c \approx 0.43$. Though a rigorous finite-size scaling is hard to obtain due to the lack of manageable system sizes, these data points shows that the central charge obviously approaches $0.5$ as the system size grows along $x$ direction. 
We remind the reader that, to make up for the lack of numerical accuracy in gapless systems, in Fig.~\ref{fig:cc} we have presented two cases: (1) a 3-legged ladder with $K_z = 0$ and $h = 0$, corresponding to the truly decoupled chains, each of which is known to be an integrable model with a $c=\frac{1}{2}$ free fermion theory, and the error in this case is truly only a finite-size error against its analytical value;   (2) the same 3-legged ladder with $K_z = 1$ and $h < h_{c2}$. If the magnetic field effectively removes the coupling along z bonds, we expect that the entropy scaling be the same as that of (1), and the error induced by the finite size in (2) should be about the same as the error induced by finite size in (1). In other words, although a precise value of $c=\frac{1}{2}$ is difficult to obtain for a strongly interacting gapless 2D system, the resemblance between (1) and (2) points out that the entanglement structure in WCC phase is the same as that of the well-known $c=1/2$ decoupled model in both scaling and finite-size induced error, i.e. the $15\%$ error due to finite size can be used also as a criteria for comparison between them, suggesting the equivalence of the two cases.

\bibliography{biblio.bib}

\end{document}

%% file: header.tex
\AtBeginDocument{
\heavyrulewidth=.08em
\lightrulewidth=.05em
\cmidrulewidth=.03em
\belowrulesep=.65ex
\belowbottomsep=0pt
\aboverulesep=.4ex
\abovetopsep=0pt
\cmidrulesep=\doublerulesep
\cmidrulekern=.5em
\defaultaddspace=.5em
}

\usepackage{float}
\usepackage{comment}
\usepackage{booktabs}
\usepackage[percent]{overpic}
\usepackage{microtype}
\usepackage{array}
\usepackage{amsmath}
\usepackage{amsthm}
\usepackage{amssymb}
\usepackage{physics}
\usepackage{graphicx}

\theoremstyle{remark}

\theoremstyle{definition}

\usepackage{bbold}
\numberwithin{thm}{section}
\usepackage{hyperref}
\usepackage[usenames,dvipsnames]{xcolor}
\usepackage{tikz}
\usetikzlibrary{arrows.meta, automata, positioning, quotes, shapes}

\bibpunct{[}{]}{,}{n}{}{}
\usepackage{hyperref} 
\hypersetup{colorlinks=true,citecolor=Red,linkcolor=Blue,urlcolor=Blue}

\newcommand{\bcen}{\begin{center}}
\newcommand{\ecen}{\end{center}}
\newcommand{\btab}{\begin{tabular}}
\newcommand{\etab}{\end{tabular}}
\newcommand{\bdes}{\begin{description}}
\newcommand{\edes}{\end{description}}

\newcommand{\beq}{\begin{equation}}
\newcommand{\eeq}{\end{equation}}
\newcommand{\bea}{\begin{eqnarray}}
\newcommand{\eea}{\end{eqnarray}}

\newcommand{\bary}{\begin{array}}
\newcommand{\eary}{\end{array}}
\newcommand{\benum}{\begin{enumerate}}
\newcommand{\eenum}{\end{enumerate}}
\newcommand{\bitem}{\begin{itemize}}
\newcommand{\eitem}{\end{itemize}}

\newcommand{\Fig}[1]{Fig.~\ref{#1}}

\makeatletter

\newcommand{\Rmnum}[1]{\expandafter\@slowromancap\romannumeral #1@}
\makeatother

\definecolor{ForestGreen}{HTML}{668000}
\definecolor{red1}{HTML}{FF4136}
\definecolor{green1}{HTML}{00802b}

\newcommand{\up}{\uparrow}
\newcommand{\dn}{\downarrow}


%% file: main.bbl
\begin{thebibliography}{56}%
\makeatletter
\providecommand \@ifxundefined [1]{%
 \@ifx{#1\undefined}
}%
\providecommand \@ifnum [1]{%
 \ifnum #1\expandafter \@firstoftwo
 \else \expandafter \@secondoftwo
 \fi
}%
\providecommand \@ifx [1]{%
 \ifx #1\expandafter \@firstoftwo
 \else \expandafter \@secondoftwo
 \fi
}%
\providecommand \natexlab [1]{#1}%
\providecommand \enquote  [1]{``#1''}%
\providecommand \bibnamefont  [1]{#1}%
\providecommand \bibfnamefont [1]{#1}%
\providecommand \citenamefont [1]{#1}%
\providecommand \href@noop [0]{\@secondoftwo}%
\providecommand \href [0]{\begingroup \@sanitize@url \@href}%
\providecommand \@href[1]{\@@startlink{#1}\@@href}%
\providecommand \@@href[1]{\endgroup#1\@@endlink}%
\providecommand \@sanitize@url [0]{\catcode `\\12\catcode `\$12\catcode
  `\&12\catcode `\#12\catcode `\^12\catcode `\_12\catcode `\%12\relax}%
\providecommand \@@startlink[1]{}%
\providecommand \@@endlink[0]{}%
\providecommand \url  [0]{\begingroup\@sanitize@url \@url }%
\providecommand \@url [1]{\endgroup\@href {#1}{\urlprefix }}%
\providecommand \urlprefix  [0]{URL }%
\providecommand \Eprint [0]{\href }%
\providecommand \doibase [0]{https://doi.org/}%
\providecommand \selectlanguage [0]{\@gobble}%
\providecommand \bibinfo  [0]{\@secondoftwo}%
\providecommand \bibfield  [0]{\@secondoftwo}%
\providecommand \translation [1]{[#1]}%
\providecommand \BibitemOpen [0]{}%
\providecommand \bibitemStop [0]{}%
\providecommand \bibitemNoStop [0]{.\EOS\space}%
\providecommand \EOS [0]{\spacefactor3000\relax}%
\providecommand \BibitemShut  [1]{\csname bibitem#1\endcsname}%
\let\auto@bib@innerbib\@empty
\bibitem [{\citenamefont {{K}itaev}(2006)}]{kitaev2006anyons}%
  \BibitemOpen
  \bibfield  {author} {\bibinfo {author} {\bibfnamefont {A.}~\bibnamefont
  {{K}itaev}},\ }\bibfield  {title} {\bibinfo {title} {Anyons in an exactly
  solved model and beyond},\ }\href
  {https://doi.org/https://doi.org/10.1016/j.aop.2005.10.005} {\bibfield
  {journal} {\bibinfo  {journal} {Annals of Physics}\ }\textbf {\bibinfo
  {volume} {321}},\ \bibinfo {pages} {2} (\bibinfo {year} {2006})}\BibitemShut
  {NoStop}%
\bibitem [{\citenamefont {Kane}\ \emph {et~al.}(2002)\citenamefont {Kane},
  \citenamefont {Mukhopadhyay},\ and\ \citenamefont {Lubensky}}]{Kane2002}%
  \BibitemOpen
  \bibfield  {author} {\bibinfo {author} {\bibfnamefont {C.~L.}\ \bibnamefont
  {Kane}}, \bibinfo {author} {\bibfnamefont {R.}~\bibnamefont {Mukhopadhyay}},\
  and\ \bibinfo {author} {\bibfnamefont {T.~C.}\ \bibnamefont {Lubensky}},\
  }\bibfield  {title} {\bibinfo {title} {Fractional quantum hall effect in an
  array of quantum wires},\ }\href
  {https://doi.org/10.1103/PhysRevLett.88.036401} {\bibfield  {journal}
  {\bibinfo  {journal} {Phys. Rev. Lett.}\ }\textbf {\bibinfo {volume} {88}},\
  \bibinfo {pages} {036401} (\bibinfo {year} {2002})}\BibitemShut {NoStop}%
\bibitem [{\citenamefont {Teo}\ and\ \citenamefont {Kane}(2014)}]{Kane2014}%
  \BibitemOpen
  \bibfield  {author} {\bibinfo {author} {\bibfnamefont {J.~C.~Y.}\
  \bibnamefont {Teo}}\ and\ \bibinfo {author} {\bibfnamefont {C.~L.}\
  \bibnamefont {Kane}},\ }\bibfield  {title} {\bibinfo {title} {From luttinger
  liquid to non-abelian quantum hall states},\ }\href
  {https://doi.org/10.1103/PhysRevB.89.085101} {\bibfield  {journal} {\bibinfo
  {journal} {Phys. Rev. B}\ }\textbf {\bibinfo {volume} {89}},\ \bibinfo
  {pages} {085101} (\bibinfo {year} {2014})}\BibitemShut {NoStop}%
\bibitem [{\citenamefont {Li}\ \emph {et~al.}(2020)\citenamefont {Li},
  \citenamefont {Ebisu}, \citenamefont {Sahoo}, \citenamefont {Oreg},\ and\
  \citenamefont {Franz}}]{Marcel2020}%
  \BibitemOpen
  \bibfield  {author} {\bibinfo {author} {\bibfnamefont {C.}~\bibnamefont
  {Li}}, \bibinfo {author} {\bibfnamefont {H.}~\bibnamefont {Ebisu}}, \bibinfo
  {author} {\bibfnamefont {S.}~\bibnamefont {Sahoo}}, \bibinfo {author}
  {\bibfnamefont {Y.}~\bibnamefont {Oreg}},\ and\ \bibinfo {author}
  {\bibfnamefont {M.}~\bibnamefont {Franz}},\ }\bibfield  {title} {\bibinfo
  {title} {Coupled wire construction of a topological phase with chiral
  tricritical ising edge modes},\ }\href
  {https://doi.org/10.1103/PhysRevB.102.165123} {\bibfield  {journal} {\bibinfo
   {journal} {Phys. Rev. B}\ }\textbf {\bibinfo {volume} {102}},\ \bibinfo
  {pages} {165123} (\bibinfo {year} {2020})}\BibitemShut {NoStop}%
\bibitem [{\citenamefont {Huang}\ \emph {et~al.}(2017)\citenamefont {Huang},
  \citenamefont {Chen}, \citenamefont {Feiguin}, \citenamefont {Chamon},\ and\
  \citenamefont {Mudry}}]{Mudry2017}%
  \BibitemOpen
  \bibfield  {author} {\bibinfo {author} {\bibfnamefont {P.-H.}\ \bibnamefont
  {Huang}}, \bibinfo {author} {\bibfnamefont {J.-H.}\ \bibnamefont {Chen}},
  \bibinfo {author} {\bibfnamefont {A.~E.}\ \bibnamefont {Feiguin}}, \bibinfo
  {author} {\bibfnamefont {C.}~\bibnamefont {Chamon}},\ and\ \bibinfo {author}
  {\bibfnamefont {C.}~\bibnamefont {Mudry}},\ }\bibfield  {title} {\bibinfo
  {title} {Coupled spin-$\frac{1}{2}$ ladders as microscopic models for
  non-abelian chiral spin liquids},\ }\href
  {https://doi.org/10.1103/PhysRevB.95.144413} {\bibfield  {journal} {\bibinfo
  {journal} {Phys. Rev. B}\ }\textbf {\bibinfo {volume} {95}},\ \bibinfo
  {pages} {144413} (\bibinfo {year} {2017})}\BibitemShut {NoStop}%
\bibitem [{\citenamefont {Slagle}\ \emph {et~al.}(2022)\citenamefont {Slagle},
  \citenamefont {Liu}, \citenamefont {Aasen}, \citenamefont {Pichler},
  \citenamefont {Mong}, \citenamefont {Chen}, \citenamefont {Endres},\ and\
  \citenamefont {Alicea}}]{Slagle2022}%
  \BibitemOpen
  \bibfield  {author} {\bibinfo {author} {\bibfnamefont {K.}~\bibnamefont
  {Slagle}}, \bibinfo {author} {\bibfnamefont {Y.}~\bibnamefont {Liu}},
  \bibinfo {author} {\bibfnamefont {D.}~\bibnamefont {Aasen}}, \bibinfo
  {author} {\bibfnamefont {H.}~\bibnamefont {Pichler}}, \bibinfo {author}
  {\bibfnamefont {R.~S.~K.}\ \bibnamefont {Mong}}, \bibinfo {author}
  {\bibfnamefont {X.}~\bibnamefont {Chen}}, \bibinfo {author} {\bibfnamefont
  {M.}~\bibnamefont {Endres}},\ and\ \bibinfo {author} {\bibfnamefont
  {J.}~\bibnamefont {Alicea}},\ }\bibfield  {title} {\bibinfo {title} {Quantum
  spin liquids bootstrapped from ising criticality in rydberg arrays},\ }\href
  {https://doi.org/10.1103/PhysRevB.106.115122} {\bibfield  {journal} {\bibinfo
   {journal} {Phys. Rev. B}\ }\textbf {\bibinfo {volume} {106}},\ \bibinfo
  {pages} {115122} (\bibinfo {year} {2022})}\BibitemShut {NoStop}%
\bibitem [{\citenamefont {Liu}\ \emph {et~al.}(2023)\citenamefont {Liu},
  \citenamefont {Tantivasadakarn}, \citenamefont {Slagle}, \citenamefont
  {Mross},\ and\ \citenamefont {Alicea}}]{liu2023}%
  \BibitemOpen
  \bibfield  {author} {\bibinfo {author} {\bibfnamefont {Y.}~\bibnamefont
  {Liu}}, \bibinfo {author} {\bibfnamefont {N.}~\bibnamefont
  {Tantivasadakarn}}, \bibinfo {author} {\bibfnamefont {K.}~\bibnamefont
  {Slagle}}, \bibinfo {author} {\bibfnamefont {D.~F.}\ \bibnamefont {Mross}},\
  and\ \bibinfo {author} {\bibfnamefont {J.}~\bibnamefont {Alicea}},\
  }\bibfield  {title} {\bibinfo {title} {Assembling {K}itaev honeycomb spin
  liquids from arrays of one-dimensional symmetry-protected topological
  phases},\ }\href {https://doi.org/10.1103/PhysRevB.108.184406} {\bibfield
  {journal} {\bibinfo  {journal} {Phys. Rev. B}\ }\textbf {\bibinfo {volume}
  {108}},\ \bibinfo {pages} {184406} (\bibinfo {year} {2023})}\BibitemShut
  {NoStop}%
\bibitem [{\citenamefont {Sebastian}\ \emph {et~al.}(2006)\citenamefont
  {Sebastian}, \citenamefont {Harrison}, \citenamefont {Batista}, \citenamefont
  {Balicas}, \citenamefont {Jaime}, \citenamefont {Sharma}, \citenamefont
  {Kawashima},\ and\ \citenamefont {Fisher}}]{Fisher2006}%
  \BibitemOpen
  \bibfield  {author} {\bibinfo {author} {\bibfnamefont {S.~E.}\ \bibnamefont
  {Sebastian}}, \bibinfo {author} {\bibfnamefont {N.}~\bibnamefont {Harrison}},
  \bibinfo {author} {\bibfnamefont {C.~D.}\ \bibnamefont {Batista}}, \bibinfo
  {author} {\bibfnamefont {L.}~\bibnamefont {Balicas}}, \bibinfo {author}
  {\bibfnamefont {M.}~\bibnamefont {Jaime}}, \bibinfo {author} {\bibfnamefont
  {P.~A.}\ \bibnamefont {Sharma}}, \bibinfo {author} {\bibfnamefont
  {N.}~\bibnamefont {Kawashima}},\ and\ \bibinfo {author} {\bibfnamefont
  {I.~R.}\ \bibnamefont {Fisher}},\ }\bibfield  {title} {\bibinfo {title}
  {Dimensional reduction at a quantum critical point},\ }\href
  {https://doi.org/10.1038/nature04732} {\bibfield  {journal} {\bibinfo
  {journal} {Nature}\ }\textbf {\bibinfo {volume} {441}},\ \bibinfo {pages}
  {617} (\bibinfo {year} {2006})}\BibitemShut {NoStop}%
\bibitem [{\citenamefont {Batista}\ \emph {et~al.}(2007)\citenamefont
  {Batista}, \citenamefont {Schmalian}, \citenamefont {Kawashima},
  \citenamefont {Sengupta}, \citenamefont {Sebastian}, \citenamefont
  {Harrison}, \citenamefont {Jaime},\ and\ \citenamefont
  {Fisher}}]{Fisher2007}%
  \BibitemOpen
  \bibfield  {author} {\bibinfo {author} {\bibfnamefont {C.~D.}\ \bibnamefont
  {Batista}}, \bibinfo {author} {\bibfnamefont {J.}~\bibnamefont {Schmalian}},
  \bibinfo {author} {\bibfnamefont {N.}~\bibnamefont {Kawashima}}, \bibinfo
  {author} {\bibfnamefont {P.}~\bibnamefont {Sengupta}}, \bibinfo {author}
  {\bibfnamefont {S.~E.}\ \bibnamefont {Sebastian}}, \bibinfo {author}
  {\bibfnamefont {N.}~\bibnamefont {Harrison}}, \bibinfo {author}
  {\bibfnamefont {M.}~\bibnamefont {Jaime}},\ and\ \bibinfo {author}
  {\bibfnamefont {I.~R.}\ \bibnamefont {Fisher}},\ }\bibfield  {title}
  {\bibinfo {title} {Geometric frustration and dimensional reduction at a
  quantum critical point},\ }\href
  {https://doi.org/10.1103/PhysRevLett.98.257201} {\bibfield  {journal}
  {\bibinfo  {journal} {Phys. Rev. Lett.}\ }\textbf {\bibinfo {volume} {98}},\
  \bibinfo {pages} {257201} (\bibinfo {year} {2007})}\BibitemShut {NoStop}%
\bibitem [{\citenamefont {Schmalian}\ and\ \citenamefont
  {Batista}(2008)}]{Batista2008}%
  \BibitemOpen
  \bibfield  {author} {\bibinfo {author} {\bibfnamefont {J.}~\bibnamefont
  {Schmalian}}\ and\ \bibinfo {author} {\bibfnamefont {C.~D.}\ \bibnamefont
  {Batista}},\ }\bibfield  {title} {\bibinfo {title} {Emergent symmetry and
  dimensional reduction at a quantum critical point},\ }\href
  {https://doi.org/10.1103/PhysRevB.77.094406} {\bibfield  {journal} {\bibinfo
  {journal} {Phys. Rev. B}\ }\textbf {\bibinfo {volume} {77}},\ \bibinfo
  {pages} {094406} (\bibinfo {year} {2008})}\BibitemShut {NoStop}%
\bibitem [{\citenamefont {Okuma}\ \emph {et~al.}(2021)\citenamefont {Okuma},
  \citenamefont {Kofu}, \citenamefont {Asai}, \citenamefont {Avdeev},
  \citenamefont {Koda}, \citenamefont {Okabe}, \citenamefont {Hiraishi},
  \citenamefont {Takeshita}, \citenamefont {Kojima}, \citenamefont {Kadono},
  \citenamefont {Masuda}, \citenamefont {Nakajima},\ and\ \citenamefont
  {Hiroi}}]{Zenji2021}%
  \BibitemOpen
  \bibfield  {author} {\bibinfo {author} {\bibfnamefont {R.}~\bibnamefont
  {Okuma}}, \bibinfo {author} {\bibfnamefont {M.}~\bibnamefont {Kofu}},
  \bibinfo {author} {\bibfnamefont {S.}~\bibnamefont {Asai}}, \bibinfo {author}
  {\bibfnamefont {M.}~\bibnamefont {Avdeev}}, \bibinfo {author} {\bibfnamefont
  {A.}~\bibnamefont {Koda}}, \bibinfo {author} {\bibfnamefont {H.}~\bibnamefont
  {Okabe}}, \bibinfo {author} {\bibfnamefont {M.}~\bibnamefont {Hiraishi}},
  \bibinfo {author} {\bibfnamefont {S.}~\bibnamefont {Takeshita}}, \bibinfo
  {author} {\bibfnamefont {K.~M.}\ \bibnamefont {Kojima}}, \bibinfo {author}
  {\bibfnamefont {R.}~\bibnamefont {Kadono}}, \bibinfo {author} {\bibfnamefont
  {T.}~\bibnamefont {Masuda}}, \bibinfo {author} {\bibfnamefont
  {K.}~\bibnamefont {Nakajima}},\ and\ \bibinfo {author} {\bibfnamefont
  {Z.}~\bibnamefont {Hiroi}},\ }\bibfield  {title} {\bibinfo {title}
  {Dimensional reduction by geometrical frustration in a cubic antiferromagnet
  composed of tetrahedral clusters},\ }\href
  {https://doi.org/10.1038/s41467-021-24636-1} {\bibfield  {journal} {\bibinfo
  {journal} {Nature Communications}\ }\textbf {\bibinfo {volume} {12}},\
  \bibinfo {pages} {4382} (\bibinfo {year} {2021})}\BibitemShut {NoStop}%
\bibitem [{\citenamefont {Wolter}\ \emph {et~al.}(2017)\citenamefont {Wolter},
  \citenamefont {Corredor}, \citenamefont {Janssen}, \citenamefont {Nenkov},
  \citenamefont {Sch\"onecker}, \citenamefont {Do}, \citenamefont {Choi},
  \citenamefont {Albrecht}, \citenamefont {Hunger}, \citenamefont {Doert},
  \citenamefont {Vojta},\ and\ \citenamefont {B\"uchner}}]{Wolter2017}%
  \BibitemOpen
  \bibfield  {author} {\bibinfo {author} {\bibfnamefont {A.~U.~B.}\
  \bibnamefont {Wolter}}, \bibinfo {author} {\bibfnamefont {L.~T.}\
  \bibnamefont {Corredor}}, \bibinfo {author} {\bibfnamefont {L.}~\bibnamefont
  {Janssen}}, \bibinfo {author} {\bibfnamefont {K.}~\bibnamefont {Nenkov}},
  \bibinfo {author} {\bibfnamefont {S.}~\bibnamefont {Sch\"onecker}}, \bibinfo
  {author} {\bibfnamefont {S.-H.}\ \bibnamefont {Do}}, \bibinfo {author}
  {\bibfnamefont {K.-Y.}\ \bibnamefont {Choi}}, \bibinfo {author}
  {\bibfnamefont {R.}~\bibnamefont {Albrecht}}, \bibinfo {author}
  {\bibfnamefont {J.}~\bibnamefont {Hunger}}, \bibinfo {author} {\bibfnamefont
  {T.}~\bibnamefont {Doert}}, \bibinfo {author} {\bibfnamefont
  {M.}~\bibnamefont {Vojta}},\ and\ \bibinfo {author} {\bibfnamefont
  {B.}~\bibnamefont {B\"uchner}},\ }\bibfield  {title} {\bibinfo {title}
  {Field-induced quantum criticality in the {K}itaev system
  $\ensuremath{\alpha}\ensuremath{-}${R}u{C}l$_{3}$},\ }\href
  {https://doi.org/10.1103/PhysRevB.96.041405} {\bibfield  {journal} {\bibinfo
  {journal} {Phys. Rev. B}\ }\textbf {\bibinfo {volume} {96}},\ \bibinfo
  {pages} {041405} (\bibinfo {year} {2017})}\BibitemShut {NoStop}%
\bibitem [{\citenamefont {Nagai}\ \emph {et~al.}(2020)\citenamefont {Nagai},
  \citenamefont {Jinno}, \citenamefont {Yoshitake}, \citenamefont {Nasu},
  \citenamefont {Motome}, \citenamefont {Itoh},\ and\ \citenamefont
  {Shimizu}}]{Nagai2020}%
  \BibitemOpen
  \bibfield  {author} {\bibinfo {author} {\bibfnamefont {Y.}~\bibnamefont
  {Nagai}}, \bibinfo {author} {\bibfnamefont {T.}~\bibnamefont {Jinno}},
  \bibinfo {author} {\bibfnamefont {J.}~\bibnamefont {Yoshitake}}, \bibinfo
  {author} {\bibfnamefont {J.}~\bibnamefont {Nasu}}, \bibinfo {author}
  {\bibfnamefont {Y.}~\bibnamefont {Motome}}, \bibinfo {author} {\bibfnamefont
  {M.}~\bibnamefont {Itoh}},\ and\ \bibinfo {author} {\bibfnamefont
  {Y.}~\bibnamefont {Shimizu}},\ }\bibfield  {title} {\bibinfo {title}
  {Two-step gap opening across the quantum critical point in the {K}itaev
  honeycomb magnet $\alpha$-{R}u{C}l$_3$},\ }\href
  {https://doi.org/10.1103/PhysRevB.101.020414} {\bibfield  {journal} {\bibinfo
   {journal} {Phys. Rev. B}\ }\textbf {\bibinfo {volume} {101}},\ \bibinfo
  {pages} {020414} (\bibinfo {year} {2020})}\BibitemShut {NoStop}%
\bibitem [{\citenamefont {Gohlke}\ \emph {et~al.}(2018)\citenamefont {Gohlke},
  \citenamefont {Wachtel}, \citenamefont {Yamaji}, \citenamefont {Pollmann},\
  and\ \citenamefont {Kim}}]{Gohlke_PRB_2018}%
  \BibitemOpen
  \bibfield  {author} {\bibinfo {author} {\bibfnamefont {M.}~\bibnamefont
  {Gohlke}}, \bibinfo {author} {\bibfnamefont {G.}~\bibnamefont {Wachtel}},
  \bibinfo {author} {\bibfnamefont {Y.}~\bibnamefont {Yamaji}}, \bibinfo
  {author} {\bibfnamefont {F.}~\bibnamefont {Pollmann}},\ and\ \bibinfo
  {author} {\bibfnamefont {Y.~B.}\ \bibnamefont {Kim}},\ }\bibfield  {title}
  {\bibinfo {title} {Quantum spin liquid signatures in {K}itaev-like frustrated
  magnets},\ }\href {https://doi.org/10.1103/PhysRevB.97.075126} {\bibfield
  {journal} {\bibinfo  {journal} {Phys. Rev. B}\ }\textbf {\bibinfo {volume}
  {97}},\ \bibinfo {pages} {075126} (\bibinfo {year} {2018})}\BibitemShut
  {NoStop}%
\bibitem [{\citenamefont {Ronquillo}\ \emph {et~al.}(2019)\citenamefont
  {Ronquillo}, \citenamefont {Vengal},\ and\ \citenamefont
  {Trivedi}}]{David2019}%
  \BibitemOpen
  \bibfield  {author} {\bibinfo {author} {\bibfnamefont {D.~C.}\ \bibnamefont
  {Ronquillo}}, \bibinfo {author} {\bibfnamefont {A.}~\bibnamefont {Vengal}},\
  and\ \bibinfo {author} {\bibfnamefont {N.}~\bibnamefont {Trivedi}},\
  }\bibfield  {title} {\bibinfo {title} {Signatures of magnetic-field-driven
  quantum phase transitions in the entanglement entropy and spin dynamics of
  the {K}itaev honeycomb model},\ }\href
  {https://doi.org/10.1103/PhysRevB.99.140413} {\bibfield  {journal} {\bibinfo
  {journal} {Phys. Rev. B}\ }\textbf {\bibinfo {volume} {99}},\ \bibinfo
  {pages} {140413} (\bibinfo {year} {2019})}\BibitemShut {NoStop}%
\bibitem [{\citenamefont {Patel}\ and\ \citenamefont
  {Trivedi}(2019)}]{Patel12199}%
  \BibitemOpen
  \bibfield  {author} {\bibinfo {author} {\bibfnamefont {N.~D.}\ \bibnamefont
  {Patel}}\ and\ \bibinfo {author} {\bibfnamefont {N.}~\bibnamefont
  {Trivedi}},\ }\bibfield  {title} {\bibinfo {title} {Magnetic field-induced
  intermediate quantum spin liquid with a spinon fermi surface},\ }\href
  {https://doi.org/10.1073/pnas.1821406116} {\bibfield  {journal} {\bibinfo
  {journal} {Proceedings of the National Academy of Sciences}\ }\textbf
  {\bibinfo {volume} {116}},\ \bibinfo {pages} {12199} (\bibinfo {year}
  {2019})}\BibitemShut {NoStop}%
\bibitem [{\citenamefont {Hickey}\ and\ \citenamefont
  {Trebst}(2019)}]{hickey2019emergence}%
  \BibitemOpen
  \bibfield  {author} {\bibinfo {author} {\bibfnamefont {C.}~\bibnamefont
  {Hickey}}\ and\ \bibinfo {author} {\bibfnamefont {S.}~\bibnamefont
  {Trebst}},\ }\bibfield  {title} {\bibinfo {title} {Emergence of a
  field-driven {U}(1) spin liquid in the {K}itaev honeycomb model},\ }\href
  {https://doi.org/https://doi.org/10.1038/s41467-019-08459-9} {\bibfield
  {journal} {\bibinfo  {journal} {Nature communications}\ }\textbf {\bibinfo
  {volume} {10}},\ \bibinfo {pages} {1} (\bibinfo {year} {2019})}\BibitemShut
  {NoStop}%
\bibitem [{\citenamefont {Hickey}\ \emph {et~al.}(2021)\citenamefont {Hickey},
  \citenamefont {Gohlke}, \citenamefont {Berke},\ and\ \citenamefont
  {Trebst}}]{Hickey20212}%
  \BibitemOpen
  \bibfield  {author} {\bibinfo {author} {\bibfnamefont {C.}~\bibnamefont
  {Hickey}}, \bibinfo {author} {\bibfnamefont {M.}~\bibnamefont {Gohlke}},
  \bibinfo {author} {\bibfnamefont {C.}~\bibnamefont {Berke}},\ and\ \bibinfo
  {author} {\bibfnamefont {S.}~\bibnamefont {Trebst}},\ }\bibfield  {title}
  {\bibinfo {title} {Generic field-driven phenomena in {K}itaev spin liquids:
  Canted magnetism and proximate spin liquid physics},\ }\href
  {https://doi.org/10.1103/PhysRevB.103.064417} {\bibfield  {journal} {\bibinfo
   {journal} {Phys. Rev. B}\ }\textbf {\bibinfo {volume} {103}},\ \bibinfo
  {pages} {064417} (\bibinfo {year} {2021})}\BibitemShut {NoStop}%
\bibitem [{\citenamefont {Trebst}\ and\ \citenamefont
  {Hickey}(2022)}]{Hickey20221}%
  \BibitemOpen
  \bibfield  {author} {\bibinfo {author} {\bibfnamefont {S.}~\bibnamefont
  {Trebst}}\ and\ \bibinfo {author} {\bibfnamefont {C.}~\bibnamefont
  {Hickey}},\ }\bibfield  {title} {\bibinfo {title} {{K}itaev materials},\
  }\href {https://doi.org/https://doi.org/10.1016/j.physrep.2021.11.003}
  {\bibfield  {journal} {\bibinfo  {journal} {Physics Reports}\ }\textbf
  {\bibinfo {volume} {950}},\ \bibinfo {pages} {1} (\bibinfo {year} {2022})},\
  \bibinfo {note} {{K}itaev materials}\BibitemShut {NoStop}%
\bibitem [{\citenamefont {Hwang}\ \emph {et~al.}(2022)\citenamefont {Hwang},
  \citenamefont {Go}, \citenamefont {Seong}, \citenamefont {Shibauchi},\ and\
  \citenamefont {Moon}}]{Hwang2022}%
  \BibitemOpen
  \bibfield  {author} {\bibinfo {author} {\bibfnamefont {K.}~\bibnamefont
  {Hwang}}, \bibinfo {author} {\bibfnamefont {A.}~\bibnamefont {Go}}, \bibinfo
  {author} {\bibfnamefont {J.~H.}\ \bibnamefont {Seong}}, \bibinfo {author}
  {\bibfnamefont {T.}~\bibnamefont {Shibauchi}},\ and\ \bibinfo {author}
  {\bibfnamefont {E.-G.}\ \bibnamefont {Moon}},\ }\bibfield  {title} {\bibinfo
  {title} {Identification of a {K}itaev quantum spin liquid by magnetic field
  angle dependence},\ }\href {https://doi.org/10.1038/s41467-021-27943-9}
  {\bibfield  {journal} {\bibinfo  {journal} {Nature Communications}\ }\textbf
  {\bibinfo {volume} {13}},\ \bibinfo {pages} {323} (\bibinfo {year}
  {2022})}\BibitemShut {NoStop}%
\bibitem [{\citenamefont {Liang}\ \emph {et~al.}(2018)\citenamefont {Liang},
  \citenamefont {Jiang}, \citenamefont {Chen}, \citenamefont {Li},\ and\
  \citenamefont {Wang}}]{Liang2018}%
  \BibitemOpen
  \bibfield  {author} {\bibinfo {author} {\bibfnamefont {S.}~\bibnamefont
  {Liang}}, \bibinfo {author} {\bibfnamefont {M.-H.}\ \bibnamefont {Jiang}},
  \bibinfo {author} {\bibfnamefont {W.}~\bibnamefont {Chen}}, \bibinfo {author}
  {\bibfnamefont {J.-X.}\ \bibnamefont {Li}},\ and\ \bibinfo {author}
  {\bibfnamefont {Q.-H.}\ \bibnamefont {Wang}},\ }\bibfield  {title} {\bibinfo
  {title} {Intermediate gapless phase and topological phase transition of the
  {K}itaev model in a uniform magnetic field},\ }\href
  {https://doi.org/10.1103/PhysRevB.98.054433} {\bibfield  {journal} {\bibinfo
  {journal} {Phys. Rev. B}\ }\textbf {\bibinfo {volume} {98}},\ \bibinfo
  {pages} {054433} (\bibinfo {year} {2018})}\BibitemShut {NoStop}%
\bibitem [{\citenamefont {Nasu}\ \emph {et~al.}(2018)\citenamefont {Nasu},
  \citenamefont {Kato}, \citenamefont {Kamiya},\ and\ \citenamefont
  {Motome}}]{Nasu2018}%
  \BibitemOpen
  \bibfield  {author} {\bibinfo {author} {\bibfnamefont {J.}~\bibnamefont
  {Nasu}}, \bibinfo {author} {\bibfnamefont {Y.}~\bibnamefont {Kato}}, \bibinfo
  {author} {\bibfnamefont {Y.}~\bibnamefont {Kamiya}},\ and\ \bibinfo {author}
  {\bibfnamefont {Y.}~\bibnamefont {Motome}},\ }\bibfield  {title} {\bibinfo
  {title} {Successive majorana topological transitions driven by a magnetic
  field in the {K}itaev model},\ }\href
  {https://doi.org/10.1103/PhysRevB.98.060416} {\bibfield  {journal} {\bibinfo
  {journal} {Phys. Rev. B}\ }\textbf {\bibinfo {volume} {98}},\ \bibinfo
  {pages} {060416} (\bibinfo {year} {2018})}\BibitemShut {NoStop}%
\bibitem [{\citenamefont {Feng}\ \emph {et~al.}(2023)\citenamefont {Feng},
  \citenamefont {Agarwala}, \citenamefont {Bhattacharjee},\ and\ \citenamefont
  {Trivedi}}]{Feng20222}%
  \BibitemOpen
  \bibfield  {author} {\bibinfo {author} {\bibfnamefont {S.}~\bibnamefont
  {Feng}}, \bibinfo {author} {\bibfnamefont {A.}~\bibnamefont {Agarwala}},
  \bibinfo {author} {\bibfnamefont {S.}~\bibnamefont {Bhattacharjee}},\ and\
  \bibinfo {author} {\bibfnamefont {N.}~\bibnamefont {Trivedi}},\ }\bibfield
  {title} {\bibinfo {title} {Anyon dynamics in field-driven phases of the
  anisotropic {K}itaev model},\ }\href
  {https://doi.org/10.1103/PhysRevB.108.035149} {\bibfield  {journal} {\bibinfo
   {journal} {Phys. Rev. B}\ }\textbf {\bibinfo {volume} {108}},\ \bibinfo
  {pages} {035149} (\bibinfo {year} {2023})}\BibitemShut {NoStop}%
\bibitem [{\citenamefont {Fransson}\ \emph {et~al.}(2016)\citenamefont
  {Fransson}, \citenamefont {Black-Schaffer},\ and\ \citenamefont
  {Balatsky}}]{Fransson2016}%
  \BibitemOpen
  \bibfield  {author} {\bibinfo {author} {\bibfnamefont {J.}~\bibnamefont
  {Fransson}}, \bibinfo {author} {\bibfnamefont {A.~M.}\ \bibnamefont
  {Black-Schaffer}},\ and\ \bibinfo {author} {\bibfnamefont {A.~V.}\
  \bibnamefont {Balatsky}},\ }\bibfield  {title} {\bibinfo {title} {Magnon
  dirac materials},\ }\href {https://doi.org/10.1103/PhysRevB.94.075401}
  {\bibfield  {journal} {\bibinfo  {journal} {Phys. Rev. B}\ }\textbf {\bibinfo
  {volume} {94}},\ \bibinfo {pages} {075401} (\bibinfo {year}
  {2016})}\BibitemShut {NoStop}%
\bibitem [{\citenamefont {Jiang}\ \emph {et~al.}(2018)\citenamefont {Jiang},
  \citenamefont {Wang}, \citenamefont {Huang},\ and\ \citenamefont
  {Lu}}]{Yuanming2018}%
  \BibitemOpen
  \bibfield  {author} {\bibinfo {author} {\bibfnamefont {H.-C.}\ \bibnamefont
  {Jiang}}, \bibinfo {author} {\bibfnamefont {C.-Y.}\ \bibnamefont {Wang}},
  \bibinfo {author} {\bibfnamefont {B.}~\bibnamefont {Huang}},\ and\ \bibinfo
  {author} {\bibfnamefont {Y.-M.}\ \bibnamefont {Lu}},\ }\href@noop {}
  {\bibinfo {title} {Field induced quantum spin liquid with spinon fermi
  surfaces in the {K}itaev model}} (\bibinfo {year} {2018}),\ \Eprint
  {https://arxiv.org/abs/1809.08247} {arXiv:1809.08247 [cond-mat.str-el]}
  \BibitemShut {NoStop}%
\bibitem [{\citenamefont {Tikhonov}\ \emph {et~al.}(2011)\citenamefont
  {Tikhonov}, \citenamefont {Feigel'man},\ and\ \citenamefont
  {{K}itaev}}]{Kitaev2011}%
  \BibitemOpen
  \bibfield  {author} {\bibinfo {author} {\bibfnamefont {K.~S.}\ \bibnamefont
  {Tikhonov}}, \bibinfo {author} {\bibfnamefont {M.~V.}\ \bibnamefont
  {Feigel'man}},\ and\ \bibinfo {author} {\bibfnamefont {A.~Y.}\ \bibnamefont
  {{K}itaev}},\ }\bibfield  {title} {\bibinfo {title} {Power-law spin
  correlations in a perturbed spin model on a honeycomb lattice},\ }\href
  {https://doi.org/10.1103/PhysRevLett.106.067203} {\bibfield  {journal}
  {\bibinfo  {journal} {Phys. Rev. Lett.}\ }\textbf {\bibinfo {volume} {106}},\
  \bibinfo {pages} {067203} (\bibinfo {year} {2011})}\BibitemShut {NoStop}%
\bibitem [{\citenamefont {Nanda}\ \emph {et~al.}(2021)\citenamefont {Nanda},
  \citenamefont {Agarwala},\ and\ \citenamefont
  {Bhattacharjee}}]{Nanda_PRB_2021}%
  \BibitemOpen
  \bibfield  {author} {\bibinfo {author} {\bibfnamefont {A.}~\bibnamefont
  {Nanda}}, \bibinfo {author} {\bibfnamefont {A.}~\bibnamefont {Agarwala}},\
  and\ \bibinfo {author} {\bibfnamefont {S.}~\bibnamefont {Bhattacharjee}},\
  }\bibfield  {title} {\bibinfo {title} {Phases and quantum phase transitions
  in the anisotropic antiferromagnetic
  {K}itaev-{H}eisenberg-$\mathrm{\ensuremath{\Gamma}}$ magnet},\ }\href
  {https://doi.org/10.1103/PhysRevB.104.195115} {\bibfield  {journal} {\bibinfo
   {journal} {Phys. Rev. B}\ }\textbf {\bibinfo {volume} {104}},\ \bibinfo
  {pages} {195115} (\bibinfo {year} {2021})}\BibitemShut {NoStop}%
\bibitem [{\citenamefont {Fishman}\ \emph
  {et~al.}(2022{\natexlab{a}})\citenamefont {Fishman}, \citenamefont {White},\
  and\ \citenamefont {Stoudenmire}}]{itensor}%
  \BibitemOpen
  \bibfield  {author} {\bibinfo {author} {\bibfnamefont {M.}~\bibnamefont
  {Fishman}}, \bibinfo {author} {\bibfnamefont {S.~R.}\ \bibnamefont {White}},\
  and\ \bibinfo {author} {\bibfnamefont {E.~M.}\ \bibnamefont {Stoudenmire}},\
  }\bibfield  {title} {\bibinfo {title} {{The ITensor Software Library for
  Tensor Network Calculations}},\ }\href
  {https://doi.org/10.21468/SciPostPhysCodeb.4} {\bibfield  {journal} {\bibinfo
   {journal} {SciPost Phys. Codebases}\ ,\ \bibinfo {pages} {4}} (\bibinfo
  {year} {2022}{\natexlab{a}})}\BibitemShut {NoStop}%
\bibitem [{\citenamefont {Fishman}\ \emph
  {et~al.}(2022{\natexlab{b}})\citenamefont {Fishman}, \citenamefont {White},\
  and\ \citenamefont {Stoudenmire}}]{itensor-r0.3}%
  \BibitemOpen
  \bibfield  {author} {\bibinfo {author} {\bibfnamefont {M.}~\bibnamefont
  {Fishman}}, \bibinfo {author} {\bibfnamefont {S.~R.}\ \bibnamefont {White}},\
  and\ \bibinfo {author} {\bibfnamefont {E.~M.}\ \bibnamefont {Stoudenmire}},\
  }\bibfield  {title} {\bibinfo {title} {{Codebase release 0.3 for ITensor}},\
  }\href {https://doi.org/10.21468/SciPostPhysCodeb.4-r0.3} {\bibfield
  {journal} {\bibinfo  {journal} {SciPost Phys. Codebases}\ ,\ \bibinfo {pages}
  {4}} (\bibinfo {year} {2022}{\natexlab{b}})}\BibitemShut {NoStop}%
\bibitem [{\citenamefont {Yogendra}\ \emph {et~al.}(2023)\citenamefont
  {Yogendra}, \citenamefont {Das},\ and\ \citenamefont
  {Baskaran}}]{yogendra2023emergent}%
  \BibitemOpen
  \bibfield  {author} {\bibinfo {author} {\bibfnamefont {K.~B.}\ \bibnamefont
  {Yogendra}}, \bibinfo {author} {\bibfnamefont {T.}~\bibnamefont {Das}},\ and\
  \bibinfo {author} {\bibfnamefont {G.}~\bibnamefont {Baskaran}},\ }\bibfield
  {title} {\bibinfo {title} {Emergent glassiness in the disorder-free {K}itaev
  model: Density matrix renormalization group study on a one-dimensional ladder
  setting},\ }\href {https://doi.org/10.1103/PhysRevB.108.165118} {\bibfield
  {journal} {\bibinfo  {journal} {Phys. Rev. B}\ }\textbf {\bibinfo {volume}
  {108}},\ \bibinfo {pages} {165118} (\bibinfo {year} {2023})}\BibitemShut
  {NoStop}%
\bibitem [{\citenamefont {Baskaran}\ \emph {et~al.}(2007)\citenamefont
  {Baskaran}, \citenamefont {Mandal},\ and\ \citenamefont
  {Shankar}}]{Baskaran2007}%
  \BibitemOpen
  \bibfield  {author} {\bibinfo {author} {\bibfnamefont {G.}~\bibnamefont
  {Baskaran}}, \bibinfo {author} {\bibfnamefont {S.}~\bibnamefont {Mandal}},\
  and\ \bibinfo {author} {\bibfnamefont {R.}~\bibnamefont {Shankar}},\
  }\bibfield  {title} {\bibinfo {title} {Exact results for spin dynamics and
  fractionalization in the {K}itaev model},\ }\href
  {https://doi.org/10.1103/PhysRevLett.98.247201} {\bibfield  {journal}
  {\bibinfo  {journal} {Phys. Rev. Lett.}\ }\textbf {\bibinfo {volume} {98}},\
  \bibinfo {pages} {247201} (\bibinfo {year} {2007})}\BibitemShut {NoStop}%
\bibitem [{\citenamefont {Zhang}\ \emph {et~al.}(2024)\citenamefont {Zhang},
  \citenamefont {Feng}, \citenamefont {Lensky}, \citenamefont {Trivedi},\ and\
  \citenamefont {Kim}}]{zhang2023machine}%
  \BibitemOpen
  \bibfield  {author} {\bibinfo {author} {\bibfnamefont {K.}~\bibnamefont
  {Zhang}}, \bibinfo {author} {\bibfnamefont {S.}~\bibnamefont {Feng}},
  \bibinfo {author} {\bibfnamefont {Y.~D.}\ \bibnamefont {Lensky}}, \bibinfo
  {author} {\bibfnamefont {N.}~\bibnamefont {Trivedi}},\ and\ \bibinfo {author}
  {\bibfnamefont {E.-A.}\ \bibnamefont {Kim}},\ }\bibfield  {title} {\bibinfo
  {title} {Machine learning reveals features of spinon fermi surface},\ }\href
  {https://doi.org/10.1038/s42005-024-01542-8} {\bibfield  {journal} {\bibinfo
  {journal} {Communications Physics}\ }\textbf {\bibinfo {volume} {7}},\
  \bibinfo {pages} {54} (\bibinfo {year} {2024})}\BibitemShut {NoStop}%
\bibitem [{\citenamefont {Feng}\ \emph
  {et~al.}(2022{\natexlab{a}})\citenamefont {Feng}, \citenamefont {He},\ and\
  \citenamefont {Trivedi}}]{Feng2022pra}%
  \BibitemOpen
  \bibfield  {author} {\bibinfo {author} {\bibfnamefont {S.}~\bibnamefont
  {Feng}}, \bibinfo {author} {\bibfnamefont {Y.}~\bibnamefont {He}},\ and\
  \bibinfo {author} {\bibfnamefont {N.}~\bibnamefont {Trivedi}},\ }\bibfield
  {title} {\bibinfo {title} {Detection of long-range entanglement in gapped
  quantum spin liquids by local measurements},\ }\href
  {https://doi.org/10.1103/PhysRevA.106.042417} {\bibfield  {journal} {\bibinfo
   {journal} {Phys. Rev. A}\ }\textbf {\bibinfo {volume} {106}},\ \bibinfo
  {pages} {042417} (\bibinfo {year} {2022}{\natexlab{a}})}\BibitemShut
  {NoStop}%
\bibitem [{\citenamefont {Sun}\ and\ \citenamefont {Chen}(2009)}]{Sun2009}%
  \BibitemOpen
  \bibfield  {author} {\bibinfo {author} {\bibfnamefont {K.-W.}\ \bibnamefont
  {Sun}}\ and\ \bibinfo {author} {\bibfnamefont {Q.-H.}\ \bibnamefont {Chen}},\
  }\bibfield  {title} {\bibinfo {title} {Quantum phase transition of the
  one-dimensional transverse-field compass model},\ }\href
  {https://doi.org/10.1103/PhysRevB.80.174417} {\bibfield  {journal} {\bibinfo
  {journal} {Phys. Rev. B}\ }\textbf {\bibinfo {volume} {80}},\ \bibinfo
  {pages} {174417} (\bibinfo {year} {2009})}\BibitemShut {NoStop}%
\bibitem [{\citenamefont {Brzezicki}\ \emph {et~al.}(2007)\citenamefont
  {Brzezicki}, \citenamefont {Dziarmaga},\ and\ \citenamefont
  {Ole\ifmmode~\acute{s}\else \'{s}\fi{}}}]{Brzezicki2007}%
  \BibitemOpen
  \bibfield  {author} {\bibinfo {author} {\bibfnamefont {W.}~\bibnamefont
  {Brzezicki}}, \bibinfo {author} {\bibfnamefont {J.}~\bibnamefont
  {Dziarmaga}},\ and\ \bibinfo {author} {\bibfnamefont {A.~M.}\ \bibnamefont
  {Ole\ifmmode~\acute{s}\else \'{s}\fi{}}},\ }\bibfield  {title} {\bibinfo
  {title} {Quantum phase transition in the one-dimensional compass model},\
  }\href {https://doi.org/10.1103/PhysRevB.75.134415} {\bibfield  {journal}
  {\bibinfo  {journal} {Phys. Rev. B}\ }\textbf {\bibinfo {volume} {75}},\
  \bibinfo {pages} {134415} (\bibinfo {year} {2007})}\BibitemShut {NoStop}%
\bibitem [{\citenamefont {Read}\ and\ \citenamefont {Green}(2000)}]{Read2000}%
  \BibitemOpen
  \bibfield  {author} {\bibinfo {author} {\bibfnamefont {N.}~\bibnamefont
  {Read}}\ and\ \bibinfo {author} {\bibfnamefont {D.}~\bibnamefont {Green}},\
  }\bibfield  {title} {\bibinfo {title} {Paired states of fermions in two
  dimensions with breaking of parity and time-reversal symmetries and the
  fractional quantum hall effect},\ }\href
  {https://doi.org/10.1103/PhysRevB.61.10267} {\bibfield  {journal} {\bibinfo
  {journal} {Phys. Rev. B}\ }\textbf {\bibinfo {volume} {61}},\ \bibinfo
  {pages} {10267} (\bibinfo {year} {2000})}\BibitemShut {NoStop}%
\bibitem [{Note1()}]{Note1}%
  \BibitemOpen
  \bibinfo {note} {For example, a stack of decoupled 1D Ising chains can become
  an intrinsically 2D model with topological order after (inverse)
  Jordan-Wigner-type transformation, as is discussed in \cite
  {Chen2007}.}\BibitemShut {Stop}%
\bibitem [{\citenamefont {Feng}\ \emph
  {et~al.}(2022{\natexlab{b}})\citenamefont {Feng}, \citenamefont {Alvarez},\
  and\ \citenamefont {Trivedi}}]{Feng2022PRB}%
  \BibitemOpen
  \bibfield  {author} {\bibinfo {author} {\bibfnamefont {S.}~\bibnamefont
  {Feng}}, \bibinfo {author} {\bibfnamefont {G.}~\bibnamefont {Alvarez}},\ and\
  \bibinfo {author} {\bibfnamefont {N.}~\bibnamefont {Trivedi}},\ }\bibfield
  {title} {\bibinfo {title} {Gapless to gapless phase transitions in quantum
  spin chains},\ }\href {https://doi.org/10.1103/PhysRevB.105.014435}
  {\bibfield  {journal} {\bibinfo  {journal} {Phys. Rev. B}\ }\textbf {\bibinfo
  {volume} {105}},\ \bibinfo {pages} {014435} (\bibinfo {year}
  {2022}{\natexlab{b}})}\BibitemShut {NoStop}%
\bibitem [{\citenamefont {Feng}\ \emph {et~al.}(2020)\citenamefont {Feng},
  \citenamefont {Patel}, \citenamefont {Kim}, \citenamefont {Han},\ and\
  \citenamefont {Trivedi}}]{Feng2020PRB}%
  \BibitemOpen
  \bibfield  {author} {\bibinfo {author} {\bibfnamefont {S.}~\bibnamefont
  {Feng}}, \bibinfo {author} {\bibfnamefont {N.~D.}\ \bibnamefont {Patel}},
  \bibinfo {author} {\bibfnamefont {P.}~\bibnamefont {Kim}}, \bibinfo {author}
  {\bibfnamefont {J.~H.}\ \bibnamefont {Han}},\ and\ \bibinfo {author}
  {\bibfnamefont {N.}~\bibnamefont {Trivedi}},\ }\bibfield  {title} {\bibinfo
  {title} {Magnetic phase transitions in quantum spin-orbital liquids},\ }\href
  {https://doi.org/10.1103/PhysRevB.101.155112} {\bibfield  {journal} {\bibinfo
   {journal} {Phys. Rev. B}\ }\textbf {\bibinfo {volume} {101}},\ \bibinfo
  {pages} {155112} (\bibinfo {year} {2020})}\BibitemShut {NoStop}%
\bibitem [{\citenamefont {Holzhey}\ \emph {et~al.}(1994)\citenamefont
  {Holzhey}, \citenamefont {Larsen},\ and\ \citenamefont
  {Wilczek}}]{HOLZHEY1994443}%
  \BibitemOpen
  \bibfield  {author} {\bibinfo {author} {\bibfnamefont {C.}~\bibnamefont
  {Holzhey}}, \bibinfo {author} {\bibfnamefont {F.}~\bibnamefont {Larsen}},\
  and\ \bibinfo {author} {\bibfnamefont {F.}~\bibnamefont {Wilczek}},\
  }\bibfield  {title} {\bibinfo {title} {Geometric and renormalized entropy in
  conformal field theory},\ }\href
  {https://doi.org/https://doi.org/10.1016/0550-3213(94)90402-2} {\bibfield
  {journal} {\bibinfo  {journal} {Nuclear Physics B}\ }\textbf {\bibinfo
  {volume} {424}},\ \bibinfo {pages} {443} (\bibinfo {year}
  {1994})}\BibitemShut {NoStop}%
\bibitem [{\citenamefont {Vidal}\ \emph {et~al.}(2003)\citenamefont {Vidal},
  \citenamefont {Latorre}, \citenamefont {Rico},\ and\ \citenamefont
  {{K}itaev}}]{Vidal2003}%
  \BibitemOpen
  \bibfield  {author} {\bibinfo {author} {\bibfnamefont {G.}~\bibnamefont
  {Vidal}}, \bibinfo {author} {\bibfnamefont {J.~I.}\ \bibnamefont {Latorre}},
  \bibinfo {author} {\bibfnamefont {E.}~\bibnamefont {Rico}},\ and\ \bibinfo
  {author} {\bibfnamefont {A.}~\bibnamefont {{K}itaev}},\ }\bibfield  {title}
  {\bibinfo {title} {Entanglement in quantum critical phenomena},\ }\href
  {https://doi.org/10.1103/PhysRevLett.90.227902} {\bibfield  {journal}
  {\bibinfo  {journal} {Phys. Rev. Lett.}\ }\textbf {\bibinfo {volume} {90}},\
  \bibinfo {pages} {227902} (\bibinfo {year} {2003})}\BibitemShut {NoStop}%
\bibitem [{\citenamefont {Latorre}\ \emph {et~al.}(2004)\citenamefont
  {Latorre}, \citenamefont {Rico},\ and\ \citenamefont {Vidal}}]{Vidal2004}%
  \BibitemOpen
  \bibfield  {author} {\bibinfo {author} {\bibfnamefont {J.~I.}\ \bibnamefont
  {Latorre}}, \bibinfo {author} {\bibfnamefont {E.}~\bibnamefont {Rico}},\ and\
  \bibinfo {author} {\bibfnamefont {G.}~\bibnamefont {Vidal}},\ }\bibfield
  {title} {\bibinfo {title} {Ground state entanglement in quantum spin
  chains},\ }\href {https://doi.org/https://doi.org/10.26421/QIC4.1-4}
  {\bibfield  {journal} {\bibinfo  {journal} {Quantum Info. Comput.}\ }\textbf
  {\bibinfo {volume} {4}},\ \bibinfo {pages} {48–92} (\bibinfo {year}
  {2004})}\BibitemShut {NoStop}%
\bibitem [{\citenamefont {Calabrese}\ and\ \citenamefont
  {Cardy}(2004)}]{Cardy2004}%
  \BibitemOpen
  \bibfield  {author} {\bibinfo {author} {\bibfnamefont {P.}~\bibnamefont
  {Calabrese}}\ and\ \bibinfo {author} {\bibfnamefont {J.}~\bibnamefont
  {Cardy}},\ }\bibfield  {title} {\bibinfo {title} {Entanglement entropy and
  quantum field theory},\ }\href
  {https://doi.org/10.1088/1742-5468/2004/06/P06002} {\bibfield  {journal}
  {\bibinfo  {journal} {Journal of Statistical Mechanics: Theory and
  Experiment}\ }\textbf {\bibinfo {volume} {2004}},\ \bibinfo {pages} {P06002}
  (\bibinfo {year} {2004})}\BibitemShut {NoStop}%
\bibitem [{\citenamefont {Calabrese}\ and\ \citenamefont
  {Cardy}(2009)}]{Calabrese_2009}%
  \BibitemOpen
  \bibfield  {author} {\bibinfo {author} {\bibfnamefont {P.}~\bibnamefont
  {Calabrese}}\ and\ \bibinfo {author} {\bibfnamefont {J.}~\bibnamefont
  {Cardy}},\ }\bibfield  {title} {\bibinfo {title} {Entanglement entropy and
  conformal field theory},\ }\href
  {https://doi.org/10.1088/1751-8113/42/50/504005} {\bibfield  {journal}
  {\bibinfo  {journal} {Journal of Physics A: Mathematical and Theoretical}\
  }\textbf {\bibinfo {volume} {42}},\ \bibinfo {pages} {504005} (\bibinfo
  {year} {2009})}\BibitemShut {NoStop}%
\bibitem [{Note2()}]{Note2}%
  \BibitemOpen
  \bibinfo {note} {Note that the decomposition into separated gauge and
  Majorana, thus the decoupling of z bonds in the mean field picture, do not
  hold on the boundaries of the cylindrical ladder. Hence, the boundary spins
  are still entangled along z bonds despite of the decoupling in the bulk,
  resulting in a smaller entanglement between the boundary spins and the bulk
  spins due to the monogamy nature of entanglement. Therefore, in order to
  catch the decoupled nature of the bulk spin chains in a finite size DMRG we
  need to fit the $S_{\protect \rm vN}(x)$ while avoiding the boundary
  spins.}\BibitemShut {Stop}%
\bibitem [{\citenamefont {Skr\o{}vseth}\ and\ \citenamefont
  {Olaussen}(2005)}]{Olaussen2005}%
  \BibitemOpen
  \bibfield  {author} {\bibinfo {author} {\bibfnamefont {S.~O.}\ \bibnamefont
  {Skr\o{}vseth}}\ and\ \bibinfo {author} {\bibfnamefont {K.}~\bibnamefont
  {Olaussen}},\ }\bibfield  {title} {\bibinfo {title} {Entanglement used to
  identify critical systems},\ }\href
  {https://doi.org/10.1103/PhysRevA.72.022318} {\bibfield  {journal} {\bibinfo
  {journal} {Phys. Rev. A}\ }\textbf {\bibinfo {volume} {72}},\ \bibinfo
  {pages} {022318} (\bibinfo {year} {2005})}\BibitemShut {NoStop}%
\bibitem [{\citenamefont {Liu}\ \emph {et~al.}(2022)\citenamefont {Liu},
  \citenamefont {Slagle}, \citenamefont {Burch},\ and\ \citenamefont
  {Alicea}}]{Liu2022}%
  \BibitemOpen
  \bibfield  {author} {\bibinfo {author} {\bibfnamefont {Y.}~\bibnamefont
  {Liu}}, \bibinfo {author} {\bibfnamefont {K.}~\bibnamefont {Slagle}},
  \bibinfo {author} {\bibfnamefont {K.~S.}\ \bibnamefont {Burch}},\ and\
  \bibinfo {author} {\bibfnamefont {J.}~\bibnamefont {Alicea}},\ }\bibfield
  {title} {\bibinfo {title} {Dynamical anyon generation in {K}itaev honeycomb
  non-abelian spin liquids},\ }\href
  {https://doi.org/10.1103/PhysRevLett.129.037201} {\bibfield  {journal}
  {\bibinfo  {journal} {Phys. Rev. Lett.}\ }\textbf {\bibinfo {volume} {129}},\
  \bibinfo {pages} {037201} (\bibinfo {year} {2022})}\BibitemShut {NoStop}%
\bibitem [{\citenamefont {Rahmani}\ \emph
  {et~al.}(2015{\natexlab{a}})\citenamefont {Rahmani}, \citenamefont {Zhu},
  \citenamefont {Franz},\ and\ \citenamefont {Affleck}}]{Affleck2015a}%
  \BibitemOpen
  \bibfield  {author} {\bibinfo {author} {\bibfnamefont {A.}~\bibnamefont
  {Rahmani}}, \bibinfo {author} {\bibfnamefont {X.}~\bibnamefont {Zhu}},
  \bibinfo {author} {\bibfnamefont {M.}~\bibnamefont {Franz}},\ and\ \bibinfo
  {author} {\bibfnamefont {I.}~\bibnamefont {Affleck}},\ }\bibfield  {title}
  {\bibinfo {title} {Emergent supersymmetry from strongly interacting majorana
  zero modes},\ }\href {https://doi.org/10.1103/PhysRevLett.115.166401}
  {\bibfield  {journal} {\bibinfo  {journal} {Phys. Rev. Lett.}\ }\textbf
  {\bibinfo {volume} {115}},\ \bibinfo {pages} {166401} (\bibinfo {year}
  {2015}{\natexlab{a}})}\BibitemShut {NoStop}%
\bibitem [{\citenamefont {Rahmani}\ \emph
  {et~al.}(2015{\natexlab{b}})\citenamefont {Rahmani}, \citenamefont {Zhu},
  \citenamefont {Franz},\ and\ \citenamefont {Affleck}}]{Affleck2015b}%
  \BibitemOpen
  \bibfield  {author} {\bibinfo {author} {\bibfnamefont {A.}~\bibnamefont
  {Rahmani}}, \bibinfo {author} {\bibfnamefont {X.}~\bibnamefont {Zhu}},
  \bibinfo {author} {\bibfnamefont {M.}~\bibnamefont {Franz}},\ and\ \bibinfo
  {author} {\bibfnamefont {I.}~\bibnamefont {Affleck}},\ }\bibfield  {title}
  {\bibinfo {title} {Phase diagram of the interacting majorana chain model},\
  }\href {https://doi.org/10.1103/PhysRevB.92.235123} {\bibfield  {journal}
  {\bibinfo  {journal} {Phys. Rev. B}\ }\textbf {\bibinfo {volume} {92}},\
  \bibinfo {pages} {235123} (\bibinfo {year} {2015}{\natexlab{b}})}\BibitemShut
  {NoStop}%
\bibitem [{\citenamefont {Rahmani}\ and\ \citenamefont
  {Franz}(2019)}]{Rahmani_2019}%
  \BibitemOpen
  \bibfield  {author} {\bibinfo {author} {\bibfnamefont {A.}~\bibnamefont
  {Rahmani}}\ and\ \bibinfo {author} {\bibfnamefont {M.}~\bibnamefont
  {Franz}},\ }\bibfield  {title} {\bibinfo {title} {Interacting majorana
  fermions},\ }\href {https://doi.org/10.1088/1361-6633/ab28ef} {\bibfield
  {journal} {\bibinfo  {journal} {Reports on Progress in Physics}\ }\textbf
  {\bibinfo {volume} {82}},\ \bibinfo {pages} {084501} (\bibinfo {year}
  {2019})}\BibitemShut {NoStop}%
\bibitem [{\citenamefont {Wu}\ \emph {et~al.}(2018)\citenamefont {Wu},
  \citenamefont {Zhu},\ and\ \citenamefont {Si}}]{Qimiao2018}%
  \BibitemOpen
  \bibfield  {author} {\bibinfo {author} {\bibfnamefont {J.}~\bibnamefont
  {Wu}}, \bibinfo {author} {\bibfnamefont {L.}~\bibnamefont {Zhu}},\ and\
  \bibinfo {author} {\bibfnamefont {Q.}~\bibnamefont {Si}},\ }\bibfield
  {title} {\bibinfo {title} {Crossovers and critical scaling in the
  one-dimensional transverse-field ising model},\ }\href
  {https://doi.org/10.1103/PhysRevB.97.245127} {\bibfield  {journal} {\bibinfo
  {journal} {Phys. Rev. B}\ }\textbf {\bibinfo {volume} {97}},\ \bibinfo
  {pages} {245127} (\bibinfo {year} {2018})}\BibitemShut {NoStop}%
\bibitem [{\citenamefont {S\o{}rensen}\ \emph
  {et~al.}(2023{\natexlab{a}})\citenamefont {S\o{}rensen}, \citenamefont
  {Gordon}, \citenamefont {Riddell}, \citenamefont {Wang},\ and\ \citenamefont
  {Kee}}]{Kee2023}%
  \BibitemOpen
  \bibfield  {author} {\bibinfo {author} {\bibfnamefont {E.~S.}\ \bibnamefont
  {S\o{}rensen}}, \bibinfo {author} {\bibfnamefont {J.}~\bibnamefont {Gordon}},
  \bibinfo {author} {\bibfnamefont {J.}~\bibnamefont {Riddell}}, \bibinfo
  {author} {\bibfnamefont {T.}~\bibnamefont {Wang}},\ and\ \bibinfo {author}
  {\bibfnamefont {H.-Y.}\ \bibnamefont {Kee}},\ }\bibfield  {title} {\bibinfo
  {title} {Field-induced chiral soliton phase in the {K}itaev spin chain},\
  }\href {https://doi.org/10.1103/PhysRevResearch.5.L012027} {\bibfield
  {journal} {\bibinfo  {journal} {Phys. Rev. Res.}\ }\textbf {\bibinfo {volume}
  {5}},\ \bibinfo {pages} {L012027} (\bibinfo {year}
  {2023}{\natexlab{a}})}\BibitemShut {NoStop}%
\bibitem [{\citenamefont {S\o{}rensen}\ \emph
  {et~al.}(2023{\natexlab{b}})\citenamefont {S\o{}rensen}, \citenamefont
  {Riddell},\ and\ \citenamefont {Kee}}]{Kee2023b}%
  \BibitemOpen
  \bibfield  {author} {\bibinfo {author} {\bibfnamefont {E.~S.}\ \bibnamefont
  {S\o{}rensen}}, \bibinfo {author} {\bibfnamefont {J.}~\bibnamefont
  {Riddell}},\ and\ \bibinfo {author} {\bibfnamefont {H.-Y.}\ \bibnamefont
  {Kee}},\ }\bibfield  {title} {\bibinfo {title} {Islands of chiral solitons in
  integer-spin {K}itaev chains},\ }\href
  {https://doi.org/10.1103/PhysRevResearch.5.013210} {\bibfield  {journal}
  {\bibinfo  {journal} {Phys. Rev. Res.}\ }\textbf {\bibinfo {volume} {5}},\
  \bibinfo {pages} {013210} (\bibinfo {year} {2023}{\natexlab{b}})}\BibitemShut
  {NoStop}%
\bibitem [{\citenamefont {Joshi}(2018)}]{Joshi2018}%
  \BibitemOpen
  \bibfield  {author} {\bibinfo {author} {\bibfnamefont {D.~G.}\ \bibnamefont
  {Joshi}},\ }\bibfield  {title} {\bibinfo {title} {Topological excitations in
  the ferromagnetic {K}itaev-{H}eisenberg model},\ }\href
  {https://doi.org/10.1103/PhysRevB.98.060405} {\bibfield  {journal} {\bibinfo
  {journal} {Phys. Rev. B}\ }\textbf {\bibinfo {volume} {98}},\ \bibinfo
  {pages} {060405} (\bibinfo {year} {2018})}\BibitemShut {NoStop}%
\bibitem [{\citenamefont {Burnell}\ and\ \citenamefont
  {Nayak}(2011)}]{Burnell2011}%
  \BibitemOpen
  \bibfield  {author} {\bibinfo {author} {\bibfnamefont {F.~J.}\ \bibnamefont
  {Burnell}}\ and\ \bibinfo {author} {\bibfnamefont {C.}~\bibnamefont
  {Nayak}},\ }\bibfield  {title} {\bibinfo {title} {{{SU}}(2) slave fermion
  solution of the {{{K}itaev}} honeycomb lattice model},\ }\href
  {https://doi.org/10.1103/PhysRevB.84.125125} {\bibfield  {journal} {\bibinfo
  {journal} {Physical Review B}\ }\textbf {\bibinfo {volume} {84}},\ \bibinfo
  {pages} {125125} (\bibinfo {year} {2011})}\BibitemShut {NoStop}%
\bibitem [{\citenamefont {Chen}\ and\ \citenamefont {Hu}(2007)}]{Chen2007}%
  \BibitemOpen
  \bibfield  {author} {\bibinfo {author} {\bibfnamefont {H.-D.}\ \bibnamefont
  {Chen}}\ and\ \bibinfo {author} {\bibfnamefont {J.}~\bibnamefont {Hu}},\
  }\bibfield  {title} {\bibinfo {title} {Exact mapping between classical and
  topological orders in two-dimensional spin systems},\ }\href
  {https://doi.org/10.1103/PhysRevB.76.193101} {\bibfield  {journal} {\bibinfo
  {journal} {Phys. Rev. B}\ }\textbf {\bibinfo {volume} {76}},\ \bibinfo
  {pages} {193101} (\bibinfo {year} {2007})}\BibitemShut {NoStop}%
\end{thebibliography}%
